\documentclass[
showkeys,11pt,
preprint,preprintnumbers,nofootinbib,
groupedaddress,superscriptaddress,amsmath,amssymb]{revtex4}
\usepackage{graphicx}
\usepackage{dcolumn}
\usepackage{bm}
\usepackage{amssymb}
\usepackage{amsmath}
\usepackage{epsfig}    
\usepackage{color}
\usepackage{slashed}
\usepackage{hhline}
\usepackage{hyperref}
\hypersetup{colorlinks,bookmarksopen,bookmarksnumbered,citecolor=blue,
linkcolor=black,pdfstartview=FitH,urlcolor=blue}

\def\be{\begin{equation}}
\def\ee{\end{equation}}
\newcommand{\bea}{\begin{eqnarray}}
\newcommand{\eea}{\end{eqnarray}}
\newcommand{\nn}{\nonumber}

\numberwithin{equation}{section}

\preprint{LPT-Orsay-15-34}
\begin{document}

\title{Two Loop Neutrino Model with Dark Matter and Leptogenesis}
\author{Shoichi Kashiwase}
\email{shoichi@hep.s.kanazawa-u.ac.jp}
\affiliation{Kanazawa University, Institute for Theoretical Physics,
Kakuma, Kanazawa, 920-1192, Japan}
\author{Hiroshi Okada}
\email{hokada@kias.re.kr}
\affiliation{School of Physics, KIAS, Seoul 130-722, Korea}
\affiliation{Physics Division, National Center for Theoretical Sciences, Hsinchu, Taiwan 300}
\author{Yuta Orikasa}
\email{orikasa@kias.re.kr}
\affiliation{School of Physics, KIAS, Seoul 130-722, Korea}
\affiliation{Department of Physics and Astronomy, Seoul National University, Seoul 151-742, Korea}
\author{Takashi Toma}
\email{takashi.toma@th.u-psud.fr}
\affiliation{Laboratoire de Physique Th\'eorique CNRS - UMR 8627,
Universit\'e de Paris-Sud 11
F-91405 Orsay Cedex, France}

\date{\today}

\begin{abstract}
We study a two-loop induced radiative neutrino model at TeV scale with
 global $U(1)$ symmetry, in which we analyze dark matter
 and resonant leptogenesis. The model includes two kinds of dark matter
 candidates. We discuss what kind of dark
 matter can satisfy the observed relic density as well
 as the current direct detection bound, and be 
 simultaneously compatible with the leptogenesis. 
 We also discuss whether our resonant leptogenesis can be differentiated
 from the other scenarios at TeV scale or not. 
\end{abstract}
\maketitle
\newpage

\section{Introduction}
After the discovery of the Higgs boson at the LHC, the Standard Model
(SM) has been established well. However 
the SM still has to be extended in order to explain the existence of dark
matter (DM), the small neutrino
masses, the baryon asymmetry in the universe and so on. 
Radiative seesaw scenarios are renowned as one of the economical models
which simultaneously explain the existence of DM and the neutrino masses. 
Since the DM candidate is necessary to generate the neutrino masses in
this kind of models, physics between DM and neutrinos is strongly correlated
in this simple framework. For example, couplings and
mass scale of DM are related with the neutrino mass scale. 
In addition, since this kind of models can naturally include a new
 particle with TeV scale mass, radiative seesaw scenarios have good testability in near future
 experiments. 
Along this line of idea, a vast literature has recently arisen in Ref.~\cite{Ma:2006km,
Aoki:2013gzs, Dasgupta:2013cwa, Krauss:2002px, Aoki:2008av,Baek:2012ub,
 Schmidt:2012yg, Bouchand:2012dx, Aoki:2011he, Farzan:2012sa,
 Bonnet:2012kz, Kumericki:2012bf, Kumericki:2012bh, Ma:2012if,
 Gil:2012ya, Okada:2012np, Hehn:2012kz, Dev:2012sg, Kajiyama:2012xg,
 Okada:2012sp, Aoki:2010ib, Kanemura:2011vm, Lindner:2011it,
 Kanemura:2011mw, Kanemura:2012rj, Gu:2007ug, Gu:2008zf, Gustafsson,
 Kajiyama:2013zla, Kajiyama:2013rla, Hernandez:2013dta,
 Hernandez:2013hea, McDonald:2013hsa, Okada:2013iba, Baek:2013fsa,
 Ma:2014cfa, Baek:2014awa, Ahriche:2014xra, Kanemura:2011jj,
 Kanemura:2013qva, Okada:2014nsa, Kanemura:2014rpa, Chen:2014ska,
 Ahriche:2014oda, Okada:2014vla, Ahriche:2014cda, Aoki:2014cja,  
Lindner:2014oea, Ahn:2012cg, Ma:2012ez, Kajiyama:2013lja,
 Kajiyama:2013sza, Ma:2013mga, Ma:2014eka, Kanemura:2015mxa,
 Bahrami:2015mwa, Baek:2015mna, Hatanaka:2014tba, Okada:2014nea,
 Sierra:2014rxa, Okada:2014qsa, Okada:2014oda, Hernandez:2015tna,
 Hernandez:2015cra, 
Culjak:2015qja, Humbert:2015epa,  Okada:2015nga, Geng:2015sza, 
Okada:2015bxa, Geng:2015coa, Ahriche:2015wha,  Ahriche:2015lba}.

On the other hand, explaining the observed baryon asymmetry in the
universe via leptogenesis is one of the challenging issues in the framework of
radiative seesaw models with right-handed neutrinos,
since couplings related to the source of the leptogenesis are
expected to be ${\cal O}$(1) due to the requirement of the neutrino
masses. It causes the strong washout of the generated baryon
asymmetry. In order to avoid this problem, we have to take
hierarchical couplings with highly degenerated masses between the source
and the mediated fields.\footnote{Such couplings can be achieved by
making use of the experimental fact that one of three neutrino
masses can be negligible. To get the
sufficient baryon asymmetry via thermal leptogenesis in the radiative
seesaw framework, resonant leptogenesis would be only the
solution that requires the mass degeneracy between the
source and the mediated fields.} 
 For example, generating the
baryon asymmetry via resonant leptogenesis has been discussed based on
the Ma model~\cite{Kashiwase:2013uy}. 

In this paper, we study a two-loop induced radiative neutrino model at
TeV scale with a global $U(1)$ symmetry, in which we analyze DM and
resonant leptogenesis simultaneously. 
In this model, we have a scalar or a fermion DM candidate. 
We discuss which kind of DM candidate can satisfy the observed relic
density as well as the current 
direct detection bound, and can also be compatible with 
leptogenesis. Since our model has two sources of leptogenesis, 
we also show different points of our resonant leptogenesis 
from the other scenarios such as the Ma model at TeV
scale~\cite{Kashiwase:2013uy}. 

This paper is organized as follows.
In Sec.~II, we show our model including field contents and their global
$U(1)$ 
charges, Higgs potential, and neutrino masses. 
In Sec.~III, DM properties including relic density and current limit
by direct detection experiments are discussed. 
In Sec.~IV, we analyze resonant leptogenesis.
Summary and conclusions are given in Sec.~V.

\section{The Model}
\subsection{Model setup}
\label{sec:model}

\begin{table}[thbp]
\centering {\fontsize{10}{12}
\begin{tabular}{|c||c|c|c|c|c||c|c|c|c|c|}
\hline   & $L_{Li}$ & $ e_{Ri} $ & $F_{L/Rj}$ & $N_{Rj}$ & $ X_{Rj} $ & $\Phi$  & $\eta$ & $\chi^0$ & $ {\chi^0}' $ & $\Sigma$  
  \\\hhline{|=#=|=|=|=|=#=|=|=|=|=|$}
$(SU(2)_L,U(1)_Y)$ & $(\bm{2},-1/2)$ & $(\bm{1},-1)$ & $(\bm{1},0)$  & $(\bm{1},0)$ & $(\bm{1},0)$ 
&  $(\bm{2},1/2)$ & $(\bm{2},1/2)$ & $(\bm{1},0)$ & $(\bm{1},0)$  & $(\bm{1},0)$
\\\hline
$U(1) $  & $-x/2$ & $-x/2$ &  $x$  & $3x/2$ & $-3x/2$ &  $0$  & $3x/2$ & $-x/2$ &  $-5x/2$ & $x$    \\\hline
Accidental $\mathbb{Z}_2$  & $+$ & $+$ &  $-$  & $+$ & $+$ &  $+$  & $-$ & $-$ &  $-$ & $+$    \\\hline
\end{tabular}%
} \caption{Field contents and charge assignments of $SU(2)_L\times
 U(1)_Y\times U(1)$, where indices $i=1-3$ and $j=1,2,(3)$ 
 represent the generation.} 
\label{tab:1}
\end{table}

As shown in Tab.~\ref{tab:1}, 
we introduce two (or three) gauge singlet vector-like fermions
$F_{L/R}$, and gauge singlet Majorana fermions $N_R$, and
$X_R$ as new fermions. 
The number of these particles should be more than two in order to obtain
at least two non-zero neutrino mass eigenvalues. 
We also introduce an inert $SU(2)_L$ doublet scalar $\eta$, two
neutral inert singlet scalars ($\chi^0, {\chi^0}'$), and a neutral
singlet scalar $\Sigma$ as new scalars. 
We assume that  only the Higgs doublet field in the SM $\Phi$ and
the new singlet scalar $\Sigma$ have vacuum 
expectation values (VEVs), which are symbolized by
$\langle\Phi\rangle=v/\sqrt{2}$ and $\langle\Sigma\rangle=v'/\sqrt{2}$
respectively.\footnote{
The scale of $v^\prime$ should be larger than $v^\prime\sim10^{7}~\mathrm{GeV}$
for successful leptogenesis as we will see later. 
Otherwise the annihilation channel $N_1N_1,X_1X_1\to GG$ whose reaction rate is
determined by $v^\prime$ does not satisfy the out-of-equilibrium
condition at $T\sim\mathcal{O}(10)~\mathrm{TeV}$ where $T$ is the
temperature of the 
universe. Thus the baryon asymmetry 
would be washed out due to this process.}  
We impose a global $U(1)$ symmetry, under which $\Phi$ does not have a
charge in order not to couple to the Goldstone boson
(GB)~\cite{Baek:2014awa}. 
The global $U(1)$ charge $x\neq0$ is in principle an arbitrary, and
the field assignments play a crucial role in realizing our neutrino
masses at two-loop level. 
If the $U(1)$ charge $x$ is fixed to be $x=2$, we
can identify this $U(1)$ symmetry as a global $B-L$ symmetry. 
Hereafter we assume 
this global $U(1)$ symmetry to be a kind of $U(1)_{B-L}$ symmetry.
Note that while the new fermions are added as vector-like and do not
contribute to anomalies, this model is anomalous since the three chiral
fermions with $B-L=-1$ corresponding to the right-handed neutrinos are not
introduced. 
If one would like to have an anomaly free model, the anomalies can be
cancelled by introducing some pairs of new
heavy vector-like fermions~\cite{Appelquist:2002mw,
Batra:2005rh}. However this is beyond the scope of this paper. 
This model has an accidental $\mathbb{Z}_2$ symmetry which can
assure the DM stability, and the accidental $\mathbb{Z}_2$ assignments
are shown in Tab.~\ref{tab:1}.

The renormalizable Lagrangian for Yukawa sector, mass term, and scalar
potential under the charge assignments are given by
\begin{eqnarray}
\mathcal{L}_{Y}
&=&
-y_\ell \bar L_L \Phi e_R  - y_{\eta} \bar L_L\eta^\dag F_R
-y_{N\chi}\bar F_L N_R \chi^0 - y_{N\chi'}\bar F_L X_R {\chi^0}'^\dag\nonumber\\
&&
- y'_{N\chi'}\bar F^c_R N_R {\chi^0}' - y'_{N\chi} \bar F^c_R X_R {\chi^0}^\dag  
- M_{NX}\bar N^c_R X_R  - M_{F}\bar F_L F_R +\rm{H.c.}, \\ 
\mathcal{V}
&=& 
 m_\Phi^2 |\Phi|^2 + m_\eta^2 |\eta|^2 + m_{\chi}^2 |\chi^0|^2 +
 m_{\chi'}^2 |{\chi^0}'|^2   + m_{\Sigma}^2 |\Sigma|^2  \nn\\
&&+ \left[
 \lambda (\Phi^\dag  \eta) \chi^0 \Sigma^\dag + \lambda' (\Phi^\dag  \eta) {\chi^0}' \Sigma 
 +\frac{\lambda''}{2} ({\chi^0}^\dag  {\chi^0}') {\Sigma^\dag}^2 + \frac{\mu_\chi}{2}(\chi^0)^2\Sigma  + {\rm H.c.}\right]
 \nn\\
&&
  +\frac{\lambda_\Phi}{4} |\Phi|^{4}   + \frac{\lambda_\eta}{4} |\eta|^{4} 
    + \frac{\lambda_{\chi}}{4} |\chi^0|^{4}  + \frac{\lambda_{\chi'}}{4} |{\chi^0}'|^{4}     + \frac{\lambda_{\Sigma}}{4} |\Sigma|^{4} 
  + \lambda_{\Phi\eta} |\Phi|^2|\eta|^2 
  + \lambda'_{\Phi\eta} (\Phi^\dagger \eta)(\eta^\dagger \Phi)\nn\\
&& +\lambda_{\Phi\chi}  |\Phi|^2|\chi^0|^2  +\lambda_{\Phi\chi'}  |\Phi|^2|{\chi^0}'|^2  +\lambda_{\Phi\Sigma}  |\Phi|^2|\Sigma|^2
+\lambda_{\eta\chi}  |\eta|^2|\chi^0|^2  +\lambda_{\eta\chi'}  |\eta|^2|{\chi^0}'|^2  +\lambda_{\eta\Sigma}  |\eta|^2|\Sigma|^2\nn\\
&&  +\lambda_{\chi\chi'}  |\chi^0|^2|{\chi^0}'|^2  +\lambda_{\chi\Sigma}  |\chi^0|^2|\Sigma|^2 + \lambda_{\chi'\Sigma}  |{\chi^0}'|^2|\Sigma|^2
,
\label{HP}
\end{eqnarray}
where the first term in $\mathcal{L}_{Y}$ generates the SM
charged lepton masses, and we assume 
the couplings $\lambda$, $\lambda'$, $\lambda''$ and $\mu_\chi$ in the
scalar potential to be
real for simplicity.
As we will see below, these couplings become important for neutrino mass generation.

After the symmetry breaking, the scalar fields can be parametrized by 
\begin{align}
&\Phi =\left(
\begin{array}{c}
w^+\\
\frac1{\sqrt2}(v+\phi+iz)
\end{array}\right),\quad 
\eta =\left(
\begin{array}{c}
\eta^+\\
\frac1{\sqrt2}(\eta_{R}^{}+i\eta_{I}^{})
\end{array}\right),\
\Sigma=\frac{v'+\sigma}{\sqrt{2}}e^{iG/v'}
.   
\label{component}
\end{align}
where $v\approx 246$ GeV is the VEV of the SM Higgs doublet, and $w^\pm$
and $z$ are respectively the GBs 
which are absorbed by the longitudinal components of the $W$ and $Z$ bosons.
Inserting the tadpole conditions,
the resulting mass matrix of the CP even scalar $(\phi,\sigma)$ 
 is given by
\begin{equation}
m^{2} (\phi,\sigma) = \left(%
\begin{array}{cc}
  \lambda_\Phi v^2 & 2 \lambda_{\Phi\Sigma}vv' \\
  2 \lambda_{\Phi\Sigma}vv' & \lambda_{\Sigma}v'^2 \\
\end{array}%
\right) \!=\! \left(\begin{array}{cc} \cos\alpha & \sin\alpha \\ -\sin\alpha & \cos\alpha \end{array}\right)
\left(\begin{array}{cc} m^2_{h} & 0 \\ 0 & m^2_{H}  \end{array}\right)
\left(\begin{array}{cc} \cos\alpha & -\sin\alpha \\ \sin\alpha &
      \cos\alpha \end{array}\right), 
\label{eq:mass_weak0}
\end{equation}
where $h$ is the SM-like Higgs boson and $H$ is an additional CP-even
Higgs mass eigenstate. 
The gauge eigenstates $\phi$ and $\sigma$ are rewritten in terms of the mass
eigenstates $h$ and $H$ as 
\begin{equation}
\left(
\begin{array}{c}
\phi\\
\sigma
\end{array}
\right)=\left(
\begin{array}{cc}
\cos\alpha & \sin\alpha\\
-\sin\alpha & \cos\alpha
\end{array}
\right)
\left(
\begin{array}{c}
h\\H
\end{array}
\right),
\quad\text{with}\quad
\sin 2\alpha=\frac{4\lambda_{\Phi\Sigma} v v'}{m^2_H-m_h^2}.
\label{eq:mass_weak}
\end{equation}
The GB $G$ in Eq.~(\ref{component}) appears due to the spontaneous symmetry breaking of
the global $U(1)$ symmetry. 
The couplings between the GB and the particles with non-trivial
global $U(1)$ charges are given by $J^\mu\partial_{\mu}G/v^\prime$
through the global $U(1)$ current $J^\mu$. As one can see, the coupling
is suppressed by the VEV $v^\prime$.

The mass matrices of the CP even and CP odd states of the inert scalar
bosons 
$(\eta,\chi^0,{\chi^0}')_{R/I}$ are respectively given by 
\begin{align}
M_R^2 &= \left(%
\begin{array}{ccc}
m^2_{\eta}+
\frac{{(\lambda_{\Phi\eta}+\lambda_{\Phi\eta}') v^2+\lambda_{\eta\Sigma}v'^2}}{2}  & \lambda vv'/2  & \lambda' vv'/2 \\
 \lambda vv'/2  & m^2_{\chi}  + \frac{\sqrt2\mu_\chi v' + \lambda_{\Phi\chi} v^2+\lambda_{\chi\Sigma}v'^2}{2} &  \lambda'' v'^2/4  \\
\lambda' vv'/2   & \lambda'' v'^2/4  &  m^2_{\chi'} + \frac{\lambda_{\Phi\chi'} v^2+\lambda_{\chi'\Sigma}v'^2}{2} \\
\end{array}%
\right),
\label{eq:real}\\
M_I^2 &= \left(%
\begin{array}{ccc}
m^2_{\eta}+\frac{(\lambda_{\Phi\eta}+\lambda_{\Phi\eta}') v^2+\lambda_{\eta\Sigma}v'^2}{2}  & -\lambda vv'/2  & -\lambda' vv'/2 \\
-\lambda vv'/2 & m^2_{\chi}  + \frac{-\sqrt2\mu_\chi v' + \lambda_{\Phi\chi} v^2+\lambda_{\chi\Sigma}v'^2}{2} &  -\lambda'' v'^2/4  \\
 -\lambda' vv'/2   & -\lambda'' v'^2/4  &  m^2_{\chi'} + \frac{\lambda_{\Phi\chi'} v^2+\lambda_{\chi'\Sigma}v'^2}{2} \\
\end{array}%
\right),
\label{eq:imaginary}
\end{align}
where we define diagonal mass matrices $(M^{2}_d)_{R/I}\equiv
(m^2_{R_1/I_1},m^2_{R_2/I_2},m^2_{R_3/I_3})$, and their 
mixing matrices $O_{id}^{R/I}$, so that they satisfy $M_{R/I}^2\equiv
{O_{id}^{R/I}}(M^{2}_d)_{R/I} ({O^{R/I}_{dj}})^T$. 
Depending on the couplings, the lightest CP even or CP odd mass
eigenstate with the mass $m_{R_1}$ or $m_{I_1}$ can be a DM
candidate.\footnote{As we will discuss later, the CP even state is identified as DM.} 
The non-standard couplings between DM and the GB induced by the
non-self-conjugate couplings
$\lambda$, $\lambda'$, $\lambda''$ and $\mu_\chi$ may be relevant to compute the DM relic
density. This coupling can be written down as
\begin{eqnarray}
\mathcal{V}_{\text{DM-DM-G-G}}=
-\left(
\frac{\lambda}{4}\frac{v}{v'}O^R_{11}O^R_{21}
+\frac{\lambda'}{4}\frac{v}{v'}O^R_{11}O^R_{31}
+\frac{\lambda''}{2}O^R_{21}O^R_{31}
+\frac{\mu_\chi}{4\sqrt{2}v^\prime}(O^R_{21})^2
\right)\text{DM}^2G^2,
\end{eqnarray}
with the mixing matrix $O^R$.
In addition, the couplings between the CP even Higgs bosons and the GB
are also 
relevant to compute the DM relic density. 
These couplings come from the kinetic term of $\Sigma$ and can be
written as 
\begin{equation}
\mathcal{L}\supset
\left[\frac{-\sin\alpha h+\cos\alpha H}{v'}
+\frac{(-\sin\alpha h+\cos\alpha H)^2}{2v'^2}
\right]\left(\partial_{\mu}G\right)\left(\partial^{\mu}G\right).
\label{eq:higgs-gb}
\end{equation}
Finally the mass eigenvalue of the charged inert scalar $\eta^+$ is given by
\begin{equation}
m^{2}_{\eta^{\pm}} = m_{\eta}^{2}  + 
\frac{\lambda_{\Phi\eta}v^{2}+\lambda_{\eta\Sigma}v'^{2}}{2}. 
\end{equation}
In this model, the typical mass scale of these new exotic particles is
assumed to be TeV scale. 
On the other hand, we should take $v^\prime\sim10^{7}~\mathrm{GeV}$ for
successful leptogenesis. 
Therefore a certain degree of tuning among the
relevant couplings cannot be avoided. 
More specifically, demanding that the diagonal elements of the mass matrix
Eq.~(\ref{eq:mass_weak0}), (\ref{eq:real}) and (\ref{eq:imaginary}) are
TeV scale and off-diagonal elements are $10~\mathrm{GeV}$ scale to obtain
small mixings of the order of $O^{R/I}_{ij}\sim10^{-2}~(i\neq j)$, 
the order of magnitude of the couplings should roughly be
$\lambda^{\prime\prime}\sim10^{-12}$ $\lambda$,
$\lambda^\prime$, $\lambda_{\Phi\Sigma}\sim10^{-7}$ and $\lambda_{\eta\Sigma}$,
$\lambda_{\chi\Sigma}$, $\lambda_{\chi^\prime\Sigma}$, $\lambda_\Sigma\sim10^{-8}$.
Although this point may be a disadvantage of this model, 
it would be worth discussing such a new concrete model with a global $U(1)$
symmetry as an example model since all the phenomenology of the neutrino masses, the
existence of DM and the baryon asymmetry of the universe are closely correlated.

\subsection{Neutrino mass matrix}
\begin{figure}[t]
\begin{center}
\includegraphics[scale=0.7]{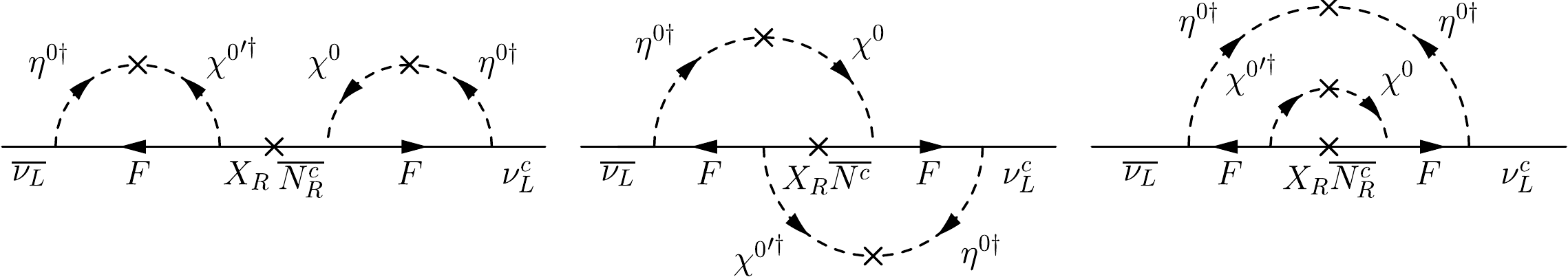}
 \caption{Radiative generation of neutrino masses.}
\label{fig:linear-seesaw}
\end{center}
\end{figure}

Due to renormalizability and the strong restriction of interactions via
the global $U(1)$
symmetry in this model, neutrino masses are not generated neither tree
level nor one-loop level. 
If the vector like fermion $F$ has a Majorana mass term, neutrino masses
would be generated at one-loop level (for example see
Ref.~\cite{Ma:2006km}), however this is not our case. 
As a result, neutrino masses are induced at two-loop level, and we have
three types of diagrams as shown in Fig.~\ref{fig:linear-seesaw}. 
The formula of the total neutrino mass matrix can be given by
\begin{eqnarray}
\left(m_\nu\right)_{\alpha\beta}
\hspace{-0.1cm}&=&\hspace{-0.1cm}
-\sum_{i,j,k}\sum_{m,n}\frac{M_{F_i}M_{F_k}}{4M_{NX_j}}\left(y_\eta\right)_{\alpha i}
\left[
\left(y_{N\chi}\right)_{ij}\left(y_{N\chi'}\right)_{kj}
+\left(y_{N\chi'}\right)_{ij}\left(y_{N\chi}\right)_{kj}
\right]
\left(y_\eta\right)_{\beta k}\nonumber\\
&&\times
\left[
I_{1(mn)}^{(ijk)}+I_{2R(mn)}^{(ijk)}+I_{3R(mn)}^{(ijk)}
\right]\nonumber\\
&&\hspace{-0.2cm}
-\sum_{i,j,k}\sum_{m,n}\frac{M_{F_i}M_{F_k}}{4M_{NX_j}}\left(y_\eta\right)_{\alpha i}
\left[
\left({y'}^*_{\!\!N\chi}\right)_{ij}\left({y'}^*_{\!\!N\chi'}\right)_{kj}
+\left({y'}^*_{\!\!N\chi'}\right)_{ij}\left({y'}^{*}_{\!\!N\chi}\right)_{kj}
\right]
\left(y_\eta\right)_{\beta k}\nonumber\\
&&\times
\left[
I_{2L(mn)}^{(ijk)}+I_{3L(mn)}^{(ijk)}
\right],
\end{eqnarray}
where $I^{(ijk)}_{1(mn)}$, a pair of $I^{(ijk)}_{2R(mn)}$ and $I^{(ijk)}_{2L(mn)}$,
a pair of $I^{(ijk)}_{3R(mn)}$ and $I^{(ijk)}_{3L(mn)}$ are the
dimensionless loop
functions which come from the left, center and right diagrams in
Fig.~\ref{fig:linear-seesaw} respectively. 
These loop functions are defined by
\begin{eqnarray}
I_{1(mn)}^{(ijk)}&=&
O^{R_mR_n}_{1213}I_{1(R_mR_n)}^{(ijk)}
-O^{R_mI_n}_{1213}I_{1(R_mI_n)}^{(ijk)}
+O^{I_mR_n}_{1213}I_{1(I_mR_n)}^{(ijk)}
-O^{I_mI_n}_{1213}I_{1(I_mI_n)}^{(ijk)},\\
I_{2R(mn)}^{(ijk)}&=&
O^{R_mR_n}_{1312}I_{2R(R_mR_n)}^{(ijk)}
+O^{R_mI_n}_{1312}I_{2R(R_mI_n)}^{(ijk)}
-O^{I_mR_n}_{1312}I_{2R(I_mR_n)}^{(ijk)}
-O^{I_mI_n}_{1312}I_{2R(I_mI_n)}^{(ijk)},\\
I_{2L(mn)}^{(ijk)}&=&
O^{R_mR_n}_{1213}I_{2L(R_mR_n)}^{(ijk)}
-O^{R_mI_n}_{1213}I_{2L(R_mI_n)}^{(ijk)}
+O^{I_mR_n}_{1213}I_{2L(I_mR_n)}^{(ijk)}
-O^{I_mI_n}_{1213}I_{2L(I_mI_n)}^{(ijk)},\\
I_{3R(mn)}^{(ijk)}&=&
O^{R_mR_n}_{1123}I_{3R(R_mR_n)}^{(ijk)}
+O^{R_mI_n}_{1123}I_{3R(R_mI_n)}^{(ijk)}
-O^{I_mR_n}_{1123}I_{3R(I_mR_n)}^{(ijk)}
-O^{I_mI_n}_{1123}I_{3R(I_mI_n)}^{(ijk)},
\label{eq:loop3R0}\\
I_{3L(mn)}^{(ijk)}&=&
O^{R_mR_n}_{1123}I_{3L(R_mR_n)}^{(ijk)}
+O^{R_mI_n}_{1123}I_{3L(R_mI_n)}^{(ijk)}
-O^{I_mR_n}_{1123}I_{3L(I_mR_n)}^{(ijk)}
-O^{I_mI_n}_{1123}I_{3L(I_mI_n)}^{(ijk)},
\label{eq:loop3L0}
\end{eqnarray}
where $O^{R_mI_n}_{abcd}=
O_{am}^{R}O_{bm}^{R}O_{cn}^{I}O_{dn}^{I}$ and 
\begin{eqnarray}
I^{(ijk)}_{1(R_mI_n)}
&=&
\frac{1}{(4\pi)^4}\left[
I\left(\frac{m_{R_m}^2}{M_{F_i}^2}\right)
I\left(\frac{m_{I_n}^2}{M_{F_k}^2}\right)
+
I\left(\frac{m_{R_m}^2}{M_{F_k}^2}\right)
I\left(\frac{m_{I_n}^2}{M_{F_i}^2}\right)
\right],\\
I^{(ijk)}_{2R(R_mI_n)}&=&
M_{NX_j}^2\!\int\!\!\frac{d^4\ell}{(2\pi)^4}\!\int\!\!\frac{d^4q}{(2\pi)^4}
\frac{1}{\ell^2-M_{F_i}^2}\frac{1}{(\ell-q)^2-M_{NX_j}^2}
\frac{1}{q^2-M_{F_k}^2}\frac{1}{\ell^2-m_{R_m}^2}
\frac{1}{q^2-m_{I_n}^2},\nonumber\\\\
I^{(ijk)}_{2L(R_mI_n)}&=&
-\frac{M_{NX_j}^2}{4M_{F_i}M_{F_k}}
\!\int\!\!\frac{d^4\ell}{(2\pi)^4}\!\int\!\!\frac{d^4q}{(2\pi)^4}
\frac{\ell\cdot q}{\ell^2-M_{F_i}^2}\frac{1}{(\ell-q)^2-M_{NX_j}^2}
\frac{1}{q^2-M_{F_k}^2}\frac{1}{\ell^2-m_{R_m}^2}
\frac{1}{q^2-m_{I_n}^2},\nonumber\\\\
I^{(ijk)}_{3R(R_mI_n)}&=&
M_{NX_j}^2\!\int\!\!\frac{d^4\ell}{(2\pi)^4}\!\int\!\!\frac{d^4q}{(2\pi)^4}
\frac{1}{\ell^2-M_{F_i}^2}\frac{1}{q^2-M_{NX_j}^2}
\frac{1}{\ell^2-M_{F_k}^2}\frac{1}{\ell^2-m_{R_m}^2}
\frac{1}{(\ell-q)^2-m_{I_n}^2},
\label{eq:loop3R}\nonumber\\\\
I^{(ijk)}_{3L(R_mI_n)}&=&
-\frac{M_{NX_j}^2}{M_{F_i}M_{F_k}}
\!\int\!\!\frac{d^4\ell}{(2\pi)^4}\!\int\!\!\frac{d^4q}{(2\pi)^4}
\frac{\ell^2}{\ell^2-M_{F_i}^2}\frac{1}{q^2-M_{NX_j}^2}
\frac{1}{\ell^2-M_{F_k}^2}\frac{1}{\ell^2-m_{R_m}^2}
\frac{1}{(\ell-q)^2-m_{I_n}^2},
\label{eq:loop3L}\nonumber\\
\end{eqnarray}
with $I(x)= x\log{x}/(1-x)$.
Note that in the derivation of the above formula, the CP phases except
the Yukawa couplings are 
neglected. 
The contribution of the left diagram can be understood as linear seesaw
like formula by splitting the diagram into two Dirac masses induced at
one-loop level. 
For the center and right diagrams, there are two kinds of contributions coming
from right and left chiralities of the internal fermions. 
In other words, these two contributions to the neutrino masses come from
the masses or momenta of the $F$ propagators in the loop respectively.
The neutrino mass generation can be understood as follows. 
Due to the global $U(1)$ symmetry breaking by the VEV
of $\Sigma$, the mixing between $\eta$, $\chi^0$
and ${\chi^0}'$ occurs. Then since the global $U(1)$ symmetry is correlated
with the lepton number conservation, the $U(1)$ symmetry breaking
implies breaking of the lepton number. 
Thus the neutrino Majorana mass term is generated after the $U(1)$
symmetry breaking. 

The neutrino mass matrix computed above can be diagonalized by the
Pontecorvo-Maki-Nakagawa-Sakata matrix $U_{\rm
PMNS}$~\cite{Maki:1962mu};
$U_{\rm PMNS}^T(m_\nu) U_{\rm PMNS}={\rm
diag}(m_{\nu_1},m_{\nu_2},m_{\nu_3})$. 
The neutrino masses and their mixing angles are measured by
experiments~\cite{Forero:2014bxa}, and these values depend on normal or
inverted mass hierarchy. 
In our model, the order of magnitude of the neutrino masses can roughly be
estimated as
$m_\nu\sim y_\eta^2Y^{2}O_{\mathrm{mix}}^4I_\mathrm{loop}v'$ where $Y$ is the
dominant Yukawa coupling in $y_{N\chi}$, $y_{N\chi'}$, $y'_{N\chi}$,
$y'_{N\chi'}$, $O_\mathrm{mix}$ represents the mixing matrix of $O^R$,
$O^I$, and $I_\mathrm{loop}$ is the loop function. 
Thus one can find that the order of $y_\eta^2Y^2\sim10^{-8}$ is required
to obtain the experimental value
$m_\nu\sim0.1~\mathrm{eV}$ with the typical assumed mixing angle
$O_\mathrm{mix}\sim10^{-2}$, $I_\mathrm{loop}\sim0.1$ and $v^\prime\sim10^{7}~\mathrm{GeV}$. 
Note that the neutrino mass matrix should be proportional to the VEV $v'$ since
$v'$ is the origin of the lepton number violation.

\begin{figure}[t]
\begin{center}
\includegraphics[scale=0.7]{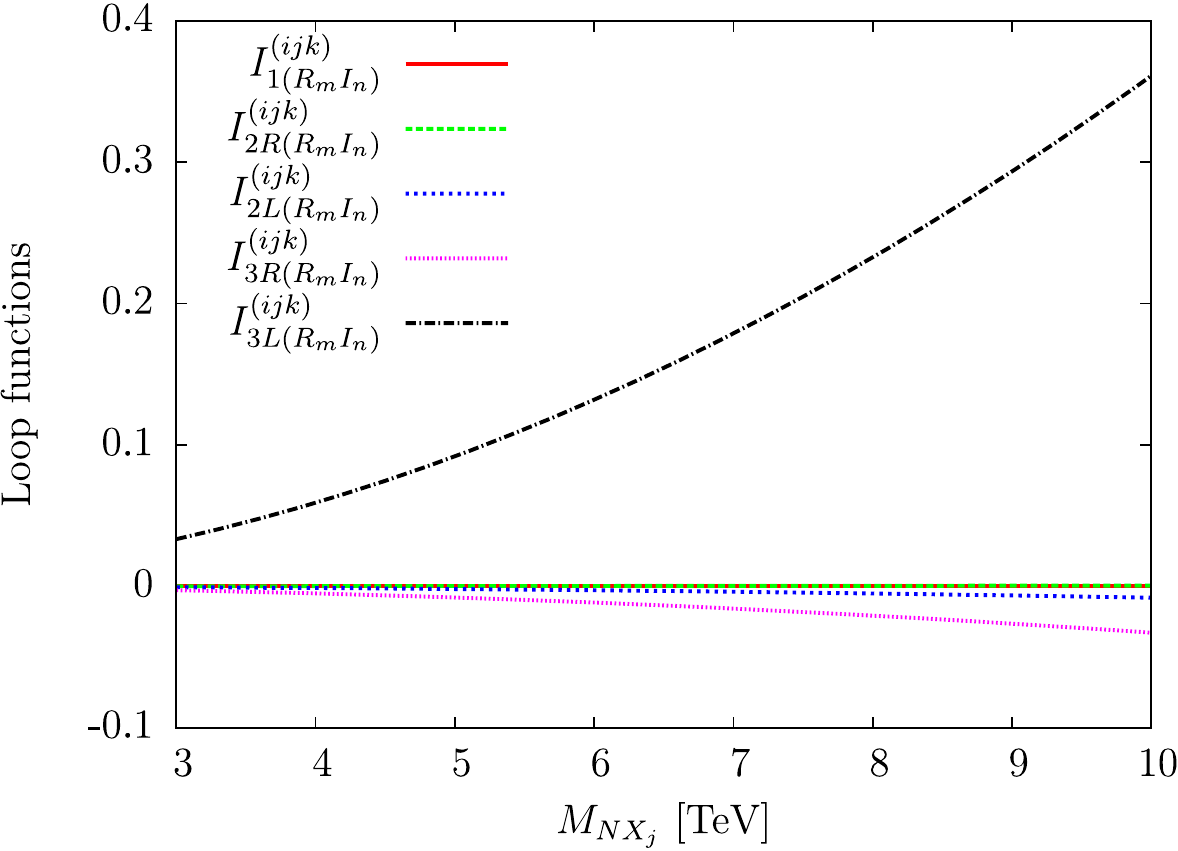}
\caption{Numerical calculation of the loop functions where the other masses are
 fixed as $M_{F_i}=M_{F_k}=1.5$ TeV and $m_{R_m}=m_{I_n}=1.2$ TeV which are
 typical sample points to discuss DM and leptogenesis as we will see below.}
\label{fig:loop}
\end{center}
\end{figure}

We numerically compute the loop functions with the public code
SecDec~\cite{Borowka:2015mxa} in order to evaluate the neutrino masses
more precisely in this model.
Here one should note that the loop functions $I_{3R(R_mI_n)}^{(ijk)}$ and
$I_{3L(R_AI_B)}^{(ijk)}$ include a divergence. This is obvious from the
definition of the loop functions in Eq.~(\ref{eq:loop3R}) and
(\ref{eq:loop3L}). 
However the divergent terms eventually cancel out with each other as
follows. 
The loop functions can be regularized with dimensional regularization
and expanded around dimension $d=4$ which can be done within
SecDec. 
With a brief evaluation, one can see that the terms including
divergences are independent on at least either of $m$ or $n$ which is the index
of the scalar mass eigenvalues. Thus the loop function including a
divergence $I_{3R/L(R_mI_n)}^{(ijk)UV}$ can be written as
$I_{3R/L(R_mI_n)}^{(ijk)UV}=I_{3R/L}^{(ijk)UV1}+I_{3R/L(R_m)}^{(ijk)UV2}+I_{3R/L(I_n)}^{(ijk)UV3}$. Taking
into account this fact and the orthogonality of the mixing matrices $O^R$
and $O^I$ which means
$\sum_{m}O^R_{im}O^R_{jm}=\sum_{m}O^I_{im}O^I_{jm}=\delta_{ij}$, one can
see that the first two terms $I_{3R/L}^{(ijk)UV1}$ and
$I_{3R/L(R_m)}^{(ijk)UV2}$ vanish after taking the summation over $n$. 
Moreover the remaining third divergent term also cancels after all the
relevant terms are summed in Eq.~(\ref{eq:loop3R0}) and (\ref{eq:loop3L0})
as  
\begin{eqnarray}
\sum_{m,n}I_{3R/L(mn)}^{(ijk)UV}
&=&
\sum_{n}
\left[
O_{2n}^{R}O_{3n}^{R}I_{3R/L(R_n)}^{(ijk)UV3}
+O_{2n}^{I}O_{3n}^{I}I_{3R/L(I_n)}^{(ijk)UV3}
-O_{2n}^{R}O_{3n}^{R}I_{3R/L(R_n)}^{(ijk)UV3}
-O_{2n}^{I}O_{3n}^{I}I_{3R/L(I_n)}^{(ijk)UV3}
\right]\nonumber\\
&=&0.
\end{eqnarray}
Thus the divergent terms do not contribute to the
neutrino masses. 
The numerical value of the loop functions are almost fixed by the maximum mass in
$M_{F_i}$, $M_{NX_j}$, $M_{F_k}$, $m_{R_m}$ and $m_{I_n}$, and the
result obtained by using SecDec is shown in
Fig.~\ref{fig:loop}.\footnote{At most $1\%$ error is included
in the numerical calculation.} 
The numerical calculation shows that the loop function
$I_{3L(R_mI_n)}^{(ijk)}$ coming from the right diagram in
Fig.~\ref{fig:linear-seesaw} 
gives a dominant contribution to the neutrino 
masses.

\subsection{LFV processes}
We should take into account 
lepton flavor violations (LFVs) such as $\mu\to e\gamma$, which
typically provide strong 
constraints on radiative neutrino mass models. 
In our case, such processes arise through only the $y_\eta$ term, and
analyses are very similar with the case of the Ma
model~\cite{Toma:2013zsa}, and the Yukawa couplings $y_{N\chi}$, $y_{N{\chi'}}$,
$y'_{N\chi}$, 
$y'_{N\chi'}$ are not constrained by the LFV processes at least
at one-loop level.
Among the LFV processes, we focus on the one-loop induced $\mu\to e
\gamma$ that gives the most stringent constraint on
$y_\eta$ and the mediating particles $F$ and $\eta^+$. 
The resulting formula for $\mu\to e\gamma$ and its experimental
bound~\cite{Adam:2013mnn} are given by 
\begin{align}
&{\rm Br}(\mu\to e\gamma)=\frac{3 \alpha_{\rm em}}{64\pi {\rm G_F^2} m_{\eta^\pm}^4}
\left|\sum_{i=1}^3(y_\eta^*)_{i1}(y_\eta)_{i2}F_2(\xi_i)\right|^2\le5.7\times 10^{-13},\label{eq:mueg}\\
&\text{with}~~~F_2(\xi_i)= \frac{1-6 \xi_i+2 \xi_i^3+3 \xi_i^2 -6 \xi_i^2 \ln\xi_i}{6(1-\xi_i)^4},
\end{align}
where ${\rm G_F}= 1.17\times 10^{-5}$ GeV$^{-2}$ is the Fermi
constant, $\xi_i=M_{F_i}^2/m_{\eta^\pm}^2$ and $\alpha_{\rm
em}= e^2/(4\pi)\approx 1/137$ is the electromagnetic fine structure
constant. 
The simplest way to avoid this constraint is to assume $y_\eta$ to be
diagonal, since its formula is proportional to the off-diagonal elements
of $y_\eta$  as one can see in Eq.~(\ref{eq:mueg}). 
In this case we expect the neutrino mixings can be derived through
the other Yukawa couplings $y_{N\chi}$, $y_{N{\chi'}}$, $y'_{N\chi}$, $y'_{N\chi'}$. 
Otherwise the parameters are constrained as $y_\eta\lesssim0.01$ and $m_{\eta^\pm}\gtrsim200~\mathrm{GeV}$. 
For instance, with the values $y_\eta=0.01$,
$m_{\eta^\pm}=200~\mathrm{GeV}$ and $F_2(\xi_i)=1/6$, the maximum
branching ratio is found to be $\mathrm{Br}(\mu\to
e\gamma)\approx1.4\times10^{-13}$. 

Since we take $y_\eta\lesssim\mathcal{O}(0.01)$ for the LFV constraint and
$\mathcal{O}(1)$ TeV of the new particle masses in the loop,
the other Yukawa
couplings $y_{N\chi}$, $y_{N\chi'}$, $y'_{N\chi}$, $y'_{N\chi'}$ should
be roughly larger than $10^{-2}$ in order to obtain the scale of
the observed neutrino mass $m_\nu\sim0.1$ eV assuming
$O_\mathrm{mix}\sim10^{-2}$ as discussed in the previous section.

\subsection{Goldstone Boson}
Here we mention some issues on the GB.
Due to the direct consequence of our global $U(1)$ symmetry, the GB remains
as a physical state, which could be constrained by some
experiments. 
In our case, the constraint comes from the invisible decay
of the SM Higgs boson, and its decay width can be computed with the coupling
given in Eq.~(\ref{eq:higgs-gb}) to be $\Gamma(h\to GG)=m_h^3
\sin^2\alpha/(32\pi v'^2)$. This decay width should be smaller than 1.2
MeV at 95$\%$ confidential level~\cite{Cheung:2013kla}, and thus we get the
constraint
\begin{equation}
\left(\frac{\sin\alpha}{0.1}\right)\left(\frac{1~\mathrm{TeV}}{v'}\right)
\lesssim2.5.
\end{equation}
Moreover, $\sin\alpha$ itself is constrained by the latest LHC searches
by ATLAS and CMS to be (conservatively)
$\sin\alpha\lesssim0.2$~\cite{Robens:2015gla}. 
Therefore the above constraint is translated to the constraint on the
VEV as $v'\gtrsim800~\mathrm{GeV}$. 
However since we take $v^\prime\sim10^7~\mathrm{GeV}$ for successful
leptogenesis, this bound is easily satisfied in our case. 

Another bound comes from the Supernova 1987A observations and
simulations, which tell us the following relation~\cite{Keung:2013mfa}: 
\begin{align}
|\lambda_{\Phi\Sigma}|\lesssim 0.011\left(\frac{m_H}{500\ {\rm MeV}}\right)^2.
\end{align}
This bound also does not affect to our model seriously, since both $m_H$
and $\lambda_{\Phi\Sigma}$ are taken to be free values of physical
parameters.

\section{Dark Matter}
\begin{figure}[t]
\begin{center}
\includegraphics[scale=0.65]{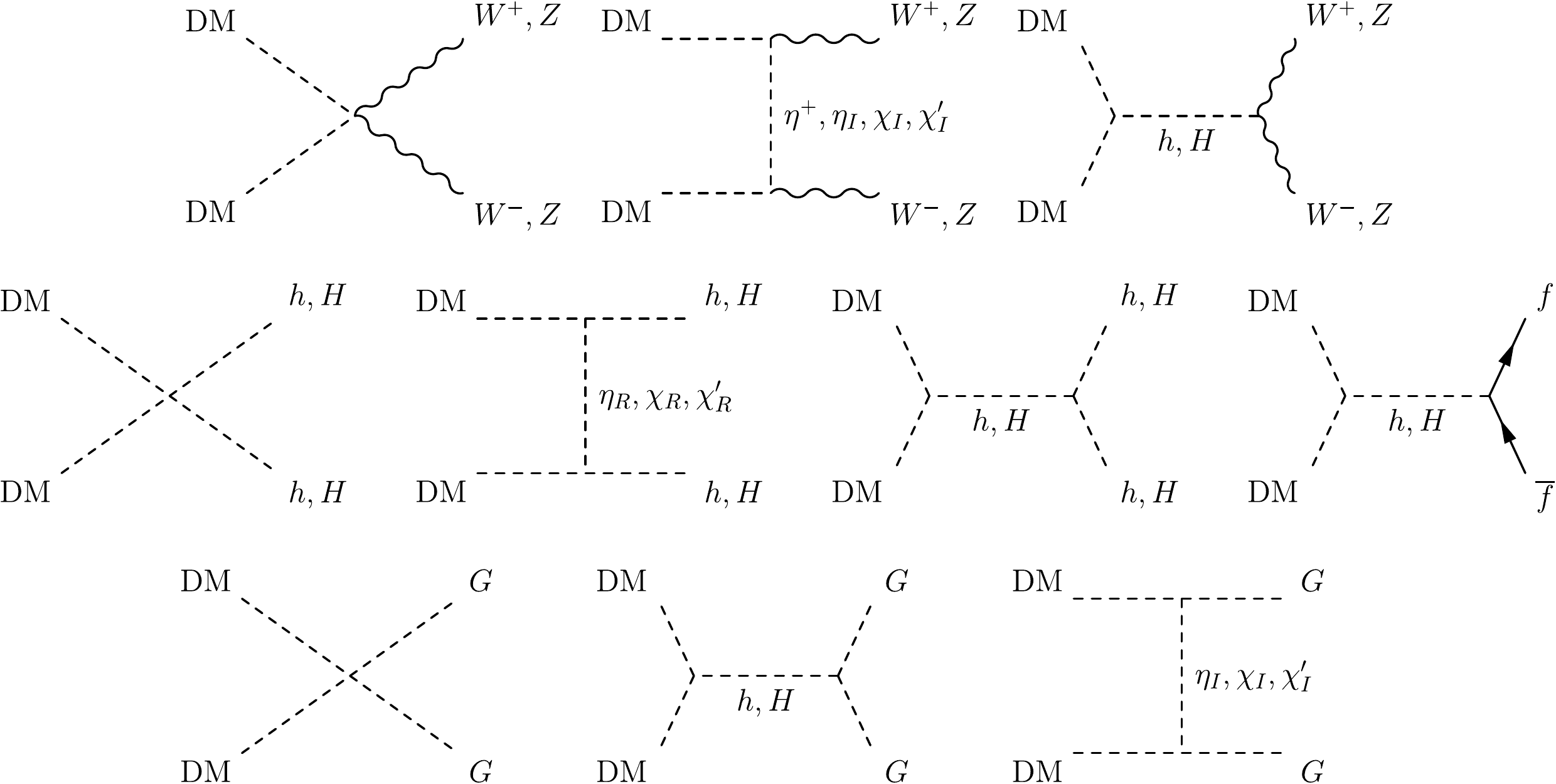}
 \caption{Diagrams of the DM annihilations where DM is identified as the
 lightest scalar mass eigenstate.}
\label{fig:DM-annihi}
\end{center}\end{figure}

We have two DM candidates which are the lightest fermion $F_1$ and the lightest mass eigenstate of
the scalars $(\eta,\chi_0,\chi'_0)_R$. 
These DM candidates can be stabilized by the accidental $\mathbb{Z}_2$
symmetry but not a remnant symmetry of the global $U(1)$ symmetry. 
This accidental $\mathbb{Z}_2$ symmetry could be understood as a kind of
the accidental symmetry which has been discussed for gauged $U(1)_{B-L}$ in Ref.~\cite{Ibe:2011hq}. 

For the fermionic DM candidate $F_1$, the relevant coupling for DM annihilations is only
the Yukawa coupling $y_\eta$, and the required strength of the Yukawa coupling is
$\mathcal{O}(1)$ for the DM mass above the electroweak scale in order to
accommodate the observed DM relic density. 
On the other hand, small coupling $y_\eta\lesssim\mathcal{O}(0.01)$
is needed to evade the LFV constraint as has been
discussed in the previous section.
Thus the fermionic DM candidate $F_1$ conflicts with
the LFV constraint.\footnote{Although the LFV constraint may be
satisfied by considering a diagonal Yukawa matrix or specific flavor
structure, we do not discuss this case.} 

Therefore we identify the lightest mass eigenstate of the scalars as a
DM candidate. 
The mixing angles among $(\eta,\chi_0,\chi'_0)_R$ are induced by the scalar couplings $\lambda$,
$\lambda'$, $\lambda''$, and the magnitude of the mixing angles should
roughly be $O_\mathrm{mix}\sim10^{-2}$ in order to reproduce the measured neutrino
masses without conflict with the $\mu\to e\gamma$ process as discussed in
the previous section. 
This order of the magnitude of the mixing angles can be achieved with
the parameter setting given in Sec.~\ref{sec:model}. 
Since the full scalar potential given by Eq.~(\ref{HP}) is rather
complicated, 
we take into account only $\lambda$, $\lambda^\prime$,
$\lambda^{\prime\prime}$, $\lambda_\Phi$, $\lambda_{\Phi\Sigma}$,
$\lambda_\Sigma$ and $\lambda_{\Phi\eta}$ for simplicity. 
Since the required order of the magnitude of the couplings $\lambda$, $\lambda^\prime$,
$\lambda^{\prime\prime}$, $\lambda_{\Phi\Sigma}$,
$\lambda_\Sigma$ is very small, they would not affect to the computation of
the DM relic density and detection probability. 
However the coupling $\lambda_{\Phi\Sigma}$ is important to induce the mixing angle
$\sin\alpha$, and $\lambda_{\Phi\eta}$ is responsible for direct
detection of DM since this coupling generates the dominant contribution
to the elastic scattering with nuclei mediated by the SM-like Higgs boson $h$.

The diagrams of the DM annihilations are shown in Fig.~\ref{fig:DM-annihi}. 
The DM couplings in the scalar potential are basically weak in our
parameter setting. However since the scalar DM candidate includes the $SU(2)_L$
doublet inert scalar $\eta_R$, DM can annihilate into the gauge bosons
via the gauge interactions in order to satisfy the observed DM relic
density if the inert doublet scalar component of the DM candidate is
sufficiently large. 
This can be achieved with a smaller $(M_R^2)_{11}$ compared to
$(M_R^2)_{22}$ and $(M_R^2)_{33}$ in Eq.~(\ref{eq:real}), and we
consider such a case. 
In this case, the annihilation channels in the first line in Fig.~\ref{fig:DM-annihi}
become dominant processes to determine the DM relic density. 

\begin{figure}[t]
\hspace{-0.2cm}
\includegraphics[scale=0.65]{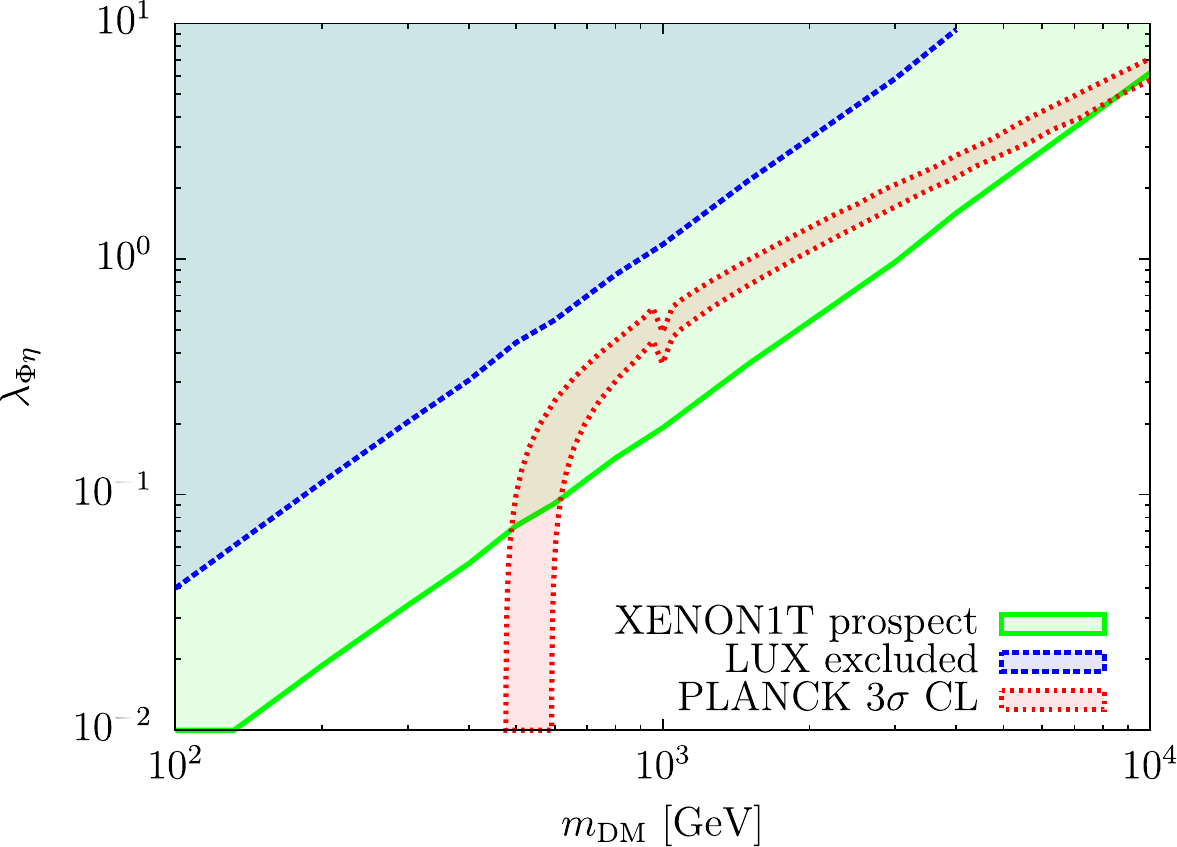}
\quad
\includegraphics[scale=0.65]{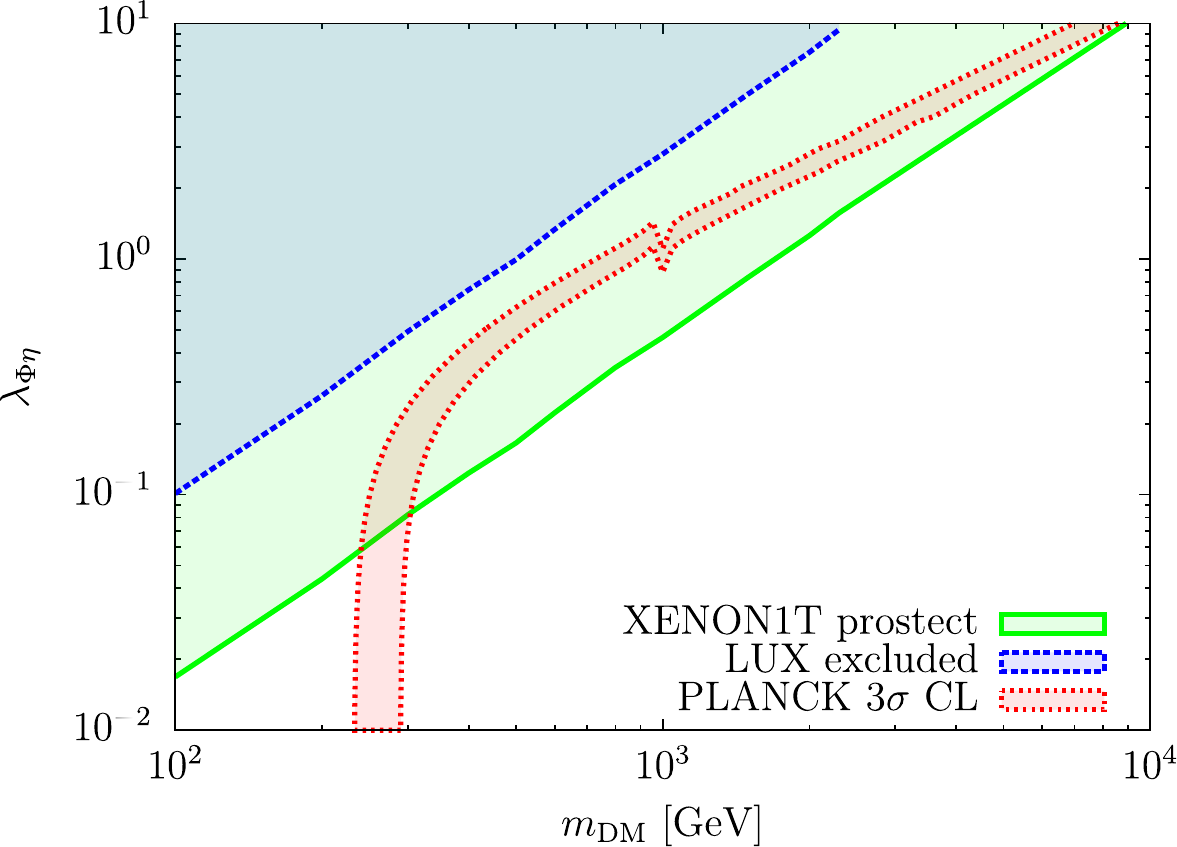}
\caption{Numerical results in the ($\lambda_{\Phi\eta}$,
 $m_\mathrm{DM}$) plane for $(\eta^0,\chi^0,{\chi^0}^\prime)_R$
 non-degenerate case (left plot) and degenerate case (right plot), where
 the parameters are fixed to be $m_H=2~\mathrm{TeV}$ and
 $\sin\alpha=0.1$.}
\label{fig:dm}
\end{figure}

The DM relic density can be evaluated by using
micrOMEGAs~\cite{Belanger:2013oya} and the
results are shown in the ($\lambda_{\Phi\eta}$, $m_\mathrm{DM}$) plane
in Fig.~\ref{fig:dm} where the heavier CP-even Higgs boson mass is taken
to be $m_H=2~\mathrm{TeV}$ and the mixing angle is $\sin\alpha=0.1$ as
an example. 
Here we take the negative values of the scalar couplings $\lambda$,
$\lambda'$, $\lambda''$. 
If the scalar couplings are positive, the lightest scalar in the
imaginary components becomes DM candidate instead of the real components.
The left plot shows the case that the masses of the other two heavier
mass eigenstates are twice of the DM mass (lightest state), and the
right plot shows the case that the heavier states are degenerate with
the DM state. 
The red colored band represents the region satisfying the observed DM
relic density by PLANCK within $3\sigma$ confidence level~\cite{Ade:2015xua}. The blue
region is excluded by the direct detection experiment LUX~\cite{Akerib:2013tjd}, and
the green region is expected to be tested by the future direct detection
experiment XENON1T~\cite{Aprile:2012zx}. 
From these plots, one can see that when the lightest DM state is
non-degenerate with the other heavier states, DM is close to the
inert doublet DM (left plot), because there is the mass threshold
$m_\mathrm{DM}\approx530~\mathrm{GeV}$ for the left plot in
Fig.~\ref{fig:dm}, which is the same property of the inert doublet
DM~\cite{Hambye:2009pw}. As well-known, the inert doublet scalar DM candidate can
satisfy the observed relic density in the mass ranges of
$m_\mathrm{DM}\sim60~\mathrm{GeV}$ and
$m_\mathrm{DM}\gtrsim530~\mathrm{GeV}$. 
On the other hand, the mass threshold can be lower as
$m_\mathrm{DM}\sim250~\mathrm{GeV}$ if $\chi^0$ and
${\chi^0}^\prime$ are degenerate with DM as one can see from the right
plot in Fig.~\ref{fig:dm}. 
This is because the interactions of the singlets $\chi^0$ and
${\chi^0}^\prime$ are described by the scalar potential, and extremely
limited in the case of the simplified potential. 
There is a small resonance feature at $m_\mathrm{DM}\sim1~\mathrm{TeV}$
due to the channel $\mathrm{DMDM}\to H^*\to \mathrm{SMSM}$, however the
resonance is not 
strong because of the small mixing angle $\sin\alpha=0.1$.
%

\section{Resonant leptogenesis}

We consider the thermal leptogenesis in this model~\cite{Fukugita:1986hr}. The lepton
number asymmetry is expected to be generated through the out-of-equilibrium decay
of the lightest Majorana fermions $N_1$ and $X_1$, if we impose the lepton
number of $F$ as $-1$. 
Although the Yukawa coupling $y_\eta$ is required to be smaller than ${\cal
O}(0.01)$ from the LFV constraint, this is large enough that $F$ and the SM leptons are in the thermal
equilibrium. 
Thus, the generated lepton number asymmetry in the $F$ sector can instantaneously be converted into the SM leptons, and then the baryon number asymmetry can be generated through sphaleron process. 

After the global $U(1)$ symmetry breaking, the Yukawa interactions for Majorana fermions
are written as 
\begin{eqnarray}
\begin{split}
{\cal L}\supset&~\bar F^c(Y_{N_i}P_L+Y_{N_i}'P_R)N_i'\chi^*+\bar F^c(Y_{X_i}P_L+Y_{X_i}'P_R)X_i'\chi^*\\
&+\bar F^c(Y_{N_i}P_L+Y_{N_i}'P_R)N_i'\chi'+\bar
 F^c(Y_{X_i}P_L+Y_{X_i}'P_R)X_i'\chi'\\ 
&+\bar F(Y_{N_i}'P_L+Y_{N_i}P_R)N_i'\chi'^*+\bar
 F(Y_{X_i}'P_L+Y_{X_i}P_R)X_i'\chi'^*\\
&+\bar F(Y_{N_i}'P_L+Y_{N_i}P_R)N_i'\chi+\bar F(Y_{X_i}'P_L+Y_{X_i}P_R)X_i'\chi,
\end{split}
\end{eqnarray}
where $N_i'$ and $X_i'$ are expressed as the mass eigenstates of each
Majorana fermion. Hereafter, we abbreviate them to
$N_i$ and $X_i$ for convenience. The Yukawa couplings are redefined as 
\begin{eqnarray}
\begin{split}
&Y_{N_i\chi}=y_{N_i\chi}^*\cos\theta_i,~Y_{N_i\chi}'=y_{N_i\chi}'\sin\theta_i,\\
&Y_{X_i\chi}=-y_{N_i\chi}^*\sin\theta_i,~Y_{X_i\chi}'=y_{X_i\chi}'\cos\theta_i,\\
&Y_{N_i\chi'}=y_{N_i\chi'}^*\sin\theta_i,~Y_{N_i\chi'}'=y'_{N_i\chi'}\cos\theta_i,\\
&Y_{X_i\chi'}=y_{X_i\chi'}^*\cos\theta_i,~Y_{X_i\chi'}'=-y_{X_i\chi'}'\sin\theta_i,\label{Ycoup}
\end{split}
\end{eqnarray}
where $\theta_i$ is the mixing angle of the $i$-th generation.

\begin{figure}[t]
\begin{center}
\includegraphics[scale=0.5]{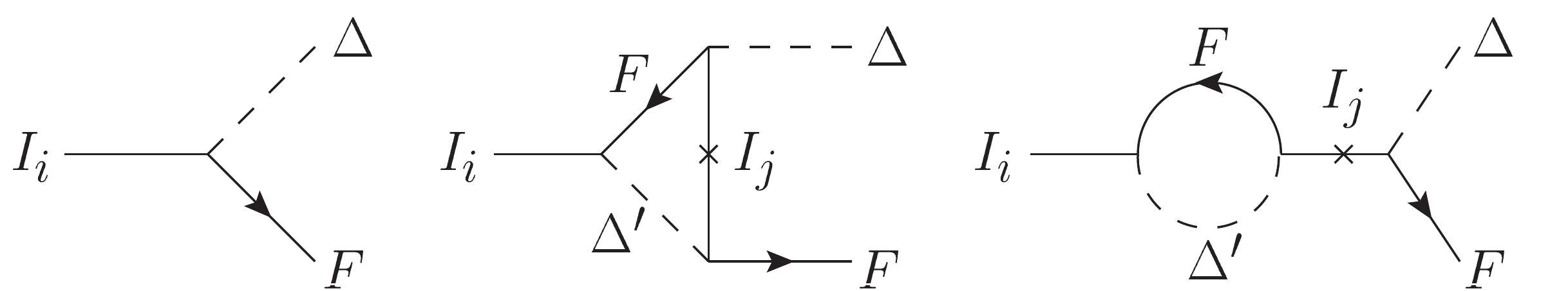}
\includegraphics[scale=0.45]{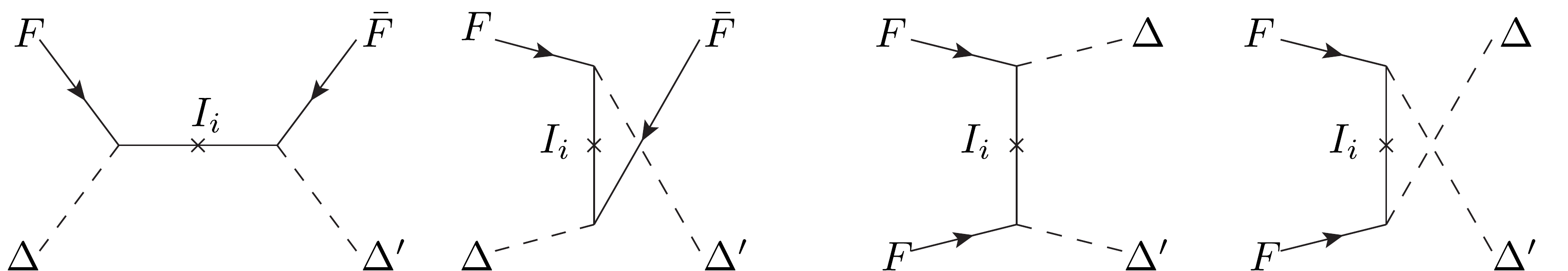}
 \caption{Lepton number violating decay and scattering processes for leptogenesis where $\Delta$ and $\Delta'$ represent the new scalar singlets, $\chi^0$ or ${\chi^0}'$ and $I_i$ ($i=1-2$) is $N_i~{\rm or}~X_i$}.
\label{fig:scattering}
\end{center}
\end{figure}
In this model, we consider that the TeV scale masses of
Majorana fermions and $v'\sim10^7~\mathrm{GeV}$ so that the annihilation
channels $N_1N_1,X_1X_1\to GG$ are decoupled from the thermal bath before
the temperature of the universe $T\sim10~\mathrm{TeV}$. 
Otherwise the generated lepton asymmetry would be washed out by these
lepton number violating processes. 

Although the TeV scale of the masses seems to be too small to realize the
sufficient $CP$ asymmetry associated with the decay processes for the
generation of the required baryon number asymmetry, the generated baryon
asymmetry can be enhanced
by the resonance effect as known in resonant leptogenesis
\cite{Flanz:1994yx, Covi:1996wh,Pilaftsis:1997jf,Akhmedov:2003dg,
Albright:2003xb,Hambye:2004jf,Pilaftsis:2003gt,Pilaftsis:2005rv, Dev:2014laa}. We
define the parameter $\epsilon$ as the amplitude of the $CP$ asymmetry.
The dominant contribution for the $CP$ asymmetry comes from the interference between
the tree diagram and the one-loop self-energy diagram as depicted in the upper line in Fig.~\ref{fig:scattering} and its formula is given by
\begin{equation}
\epsilon \propto
 \frac{(M_{I_1}^2-M_{I_2}^2)M_{I_2}\Gamma_{I_2}}{(M_{I_1}^2-M_{I_2}^2)^2+M_{I_2}^2\Gamma_{I_2}^2}, 
\end{equation}
where $M_{I_i}$ and $\Gamma_{I_i}$ are the mass and the decay rate of
$I_i$ $(I=N~{\rm or}~X)$ respectively. From this equation, the maximum
enhancement is caused when $M_{I_1}^2-M_{I_2}^2=M_{I_2}\Gamma_{I_2}$. 
In this model, we can take a larger $\Gamma_{I_2}$ compared to that of
the second lightest right-handed neutrino in the canonical resonant
leptogenesis at TeV scale, since the Yukawa couplings can be large without
conflicting with the observed neutrino masses due to
the loop suppression. 
Thus, the required magnitude of the degeneracy of the Majorana fermion mass can be quite milder
to generate the sufficient baryon number asymmetry than those in the usual
resonant leptogenesis at TeV scale
\cite{Kashiwase:2012xd,Kashiwase:2013uy}. 
{One may think that the baryon asymmetry should be correlated with the VEV
of the singlet scalar $\Sigma$ since the $B-L$ breaking occurs with only
$v'$.
Indeed this $B-L$ breaking effect is included in the total mass
matrix of $N_i$, $X_i$ and the active neutrinos $\nu_i$. 
For example, the $B-L$ violating Dirac mass term between $\nu_i$
and $X_j$ is induced at one-loop level as can be seen from the left
diagram in Fig.~\ref{fig:linear-seesaw}. Thus the effect of the 
$B-L$ breaking is included in the masses of $N_i$ and $X_i$, and 
one can understand that the baryon asymmetry is generated through the
breaking effect in the mass matrix. 
}

We need to take into account washout effects to evaluate the baryon
number asymmetry. The generated lepton number asymmetry could be washed
out through the lepton number violating 2-2 scattering processes and the
inverse decay of $I_i$. However, if the relevant Yukawa couplings are
small enough, these processes can be nearly decoupled before the
temperature of the thermal plasma decreases to
$T\hspace{0.3em}\raisebox{0.4ex}{$<$}\hspace{-0.75em}\raisebox{-.7ex}{$\sim$}\hspace{0.3em}M_{I_1}$. Thus,
the washout of the generated lepton number asymmetry is expected to be
suppressed sufficiently in this period. In order to examine this quantitatively, we
numerically solve the coupled Boltzmann equations for the number density
of $N_1,~X_1$ and the lepton number asymmetry. We introduce the number
density of $N_1$, $X_1$ and the lepton number asymmetry in the comoving
volume as $Y_{N_1}=n_{N_1}/s$, $Y_{X_1}=n_{X_1}/s$ and
$Y_{F} =(n_F-n_{\bar F})/s$ respectively, by using the entropy
density $s$ and the number densities which are expressed by
$n_{N_1},~n_{X_1}$, $n_{F}$ and $n_{\bar{F}}$. Their equilibrium values
are given by $Y_{I_1}^{\rm eq}=\frac{45}{2\pi^4g_*}z^2K_2(z)$, where $z$ is
defined by $z=M_{I_1}/T$, $g_*$ is the number of relativistic
degrees of freedom and $K_2(z)$ is the modified Bessel function of the
second kind with the order $2$. Since we assume $Y_F$ is immediately translated into the SM
leptons as we mentioned above, we use the relation $B
=\frac{8}{23}(B-L)$ which is derived from the chemical equilibrium
condition in this model, and the baryon number asymmetry $Y_{B}$ in the
present Universe is estimated as $Y_B=-\frac{8}{23}Y_F(z_{\rm EW})$,
where $z_{\rm EW}=M_{I_1}/T_{\rm EW}$ is related to the sphaleron decoupling temperature
$T_{\rm EW}$. 

The coupled Boltzmann equations for the leptogenesis in our model are written as~\cite{Kolb:1979qa}
\begin{eqnarray}
&&\frac{dY_{N_1}}{dz}=-\frac{z}{sH}\left(\frac{Y_{N_1}}{Y_{N_1}^{eq}}-1\right)\gamma_{N_1}^D,\\
&&\frac{dY_{X_1}}{dz}=-\frac{z}{sH}\left(\frac{Y_{X_1}}{Y_{X_1}^{eq}}-1\right)\gamma_{X_1}^D,\\
\nonumber&&
\frac{dY_F}{dz}=\frac{z}{sH}\Biggl\{\epsilon_N\left(\frac{Y_{N_1}}{Y_{N_1}^{eq}}-1\right)\gamma_{N_1}^D+\epsilon_X
\left(\frac{Y_{X_1}}{Y_{X_1}^{eq}}-1\right)\gamma_{X_1}^D\\
&&\qquad\quad-\frac{2Y_F}{Y_F^{eq}}\Biggl[\sum_i\frac{\gamma_{N_i}+\gamma_{X_i}}{4}+\sum_{\Delta,\Delta'}(\gamma_{F\Delta
 F\Delta'}+\gamma_{FF\Delta\Delta'})\Biggr]\Biggr\}, 
\end{eqnarray}
where $\gamma_{I_i}^D$ is defined by
\begin{equation}
\gamma_{I_i}^D=\frac{M_{I_i}^2T}{\pi^2} K_1\left(\frac{M_{I_i}}{T}\right)\Gamma_{I_i}^D,
\end{equation}
with the modified Bessel function of the second kind $K_1(z)$ with the
order $1$,
$\gamma_{abij}$ is the reaction density for the scattering process
$ab\leftrightarrow ij$ which is given by 
\begin{equation}
\gamma_{abij}=\frac{T}{64\pi^4}\int^\infty_{s_{\rm min}}ds~\hat\sigma_{abij}(s)\sqrt{s} K_1\left(\frac{\sqrt{s}}{T}\right),
\end{equation}
with $s_{\rm min}={\rm max}\left[(m_a+m_b)^2,(m_i+m_j)^2\right]$ and the
reduced cross section $\hat\sigma_{abij}(s)$. 
There are two kinds of the processes $F\Delta\leftrightarrow
\bar{F}\Delta'$ and $FF\leftrightarrow\Delta\Delta'$
for the scattering processes as depicted in the bottom line of
Fig.~\ref{fig:scattering}. 
The reduced cross section in our model is rather complicated since a lot
of particles exist, but can be straightforwardly computed
from the Lagrangian as same as Ref.~\cite{Luty:1992un, Plumacher:1997ru}. We solve the coupled
Boltzmann equations numerically. 

\begin{figure}[t]
\includegraphics[scale=0.6]{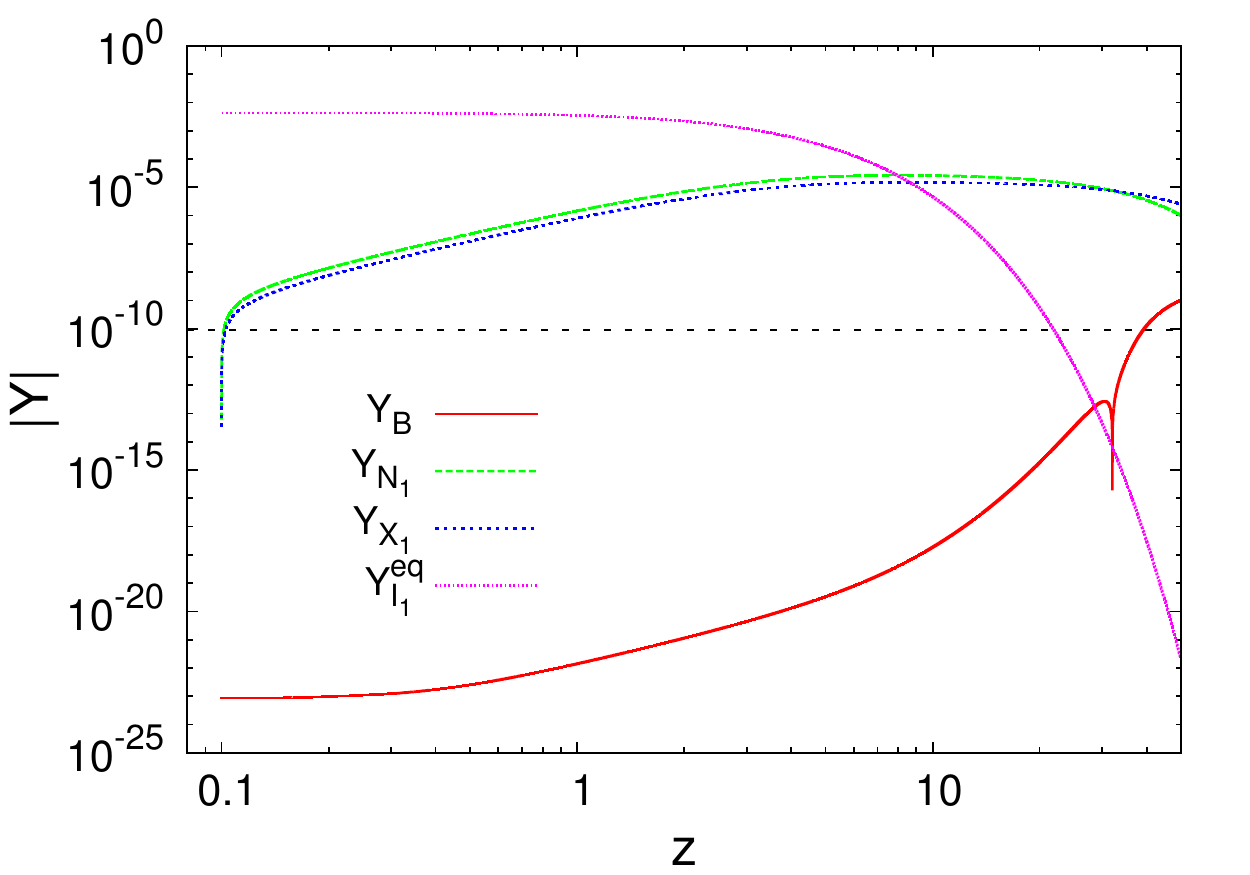}
\quad
\includegraphics[scale=0.6]{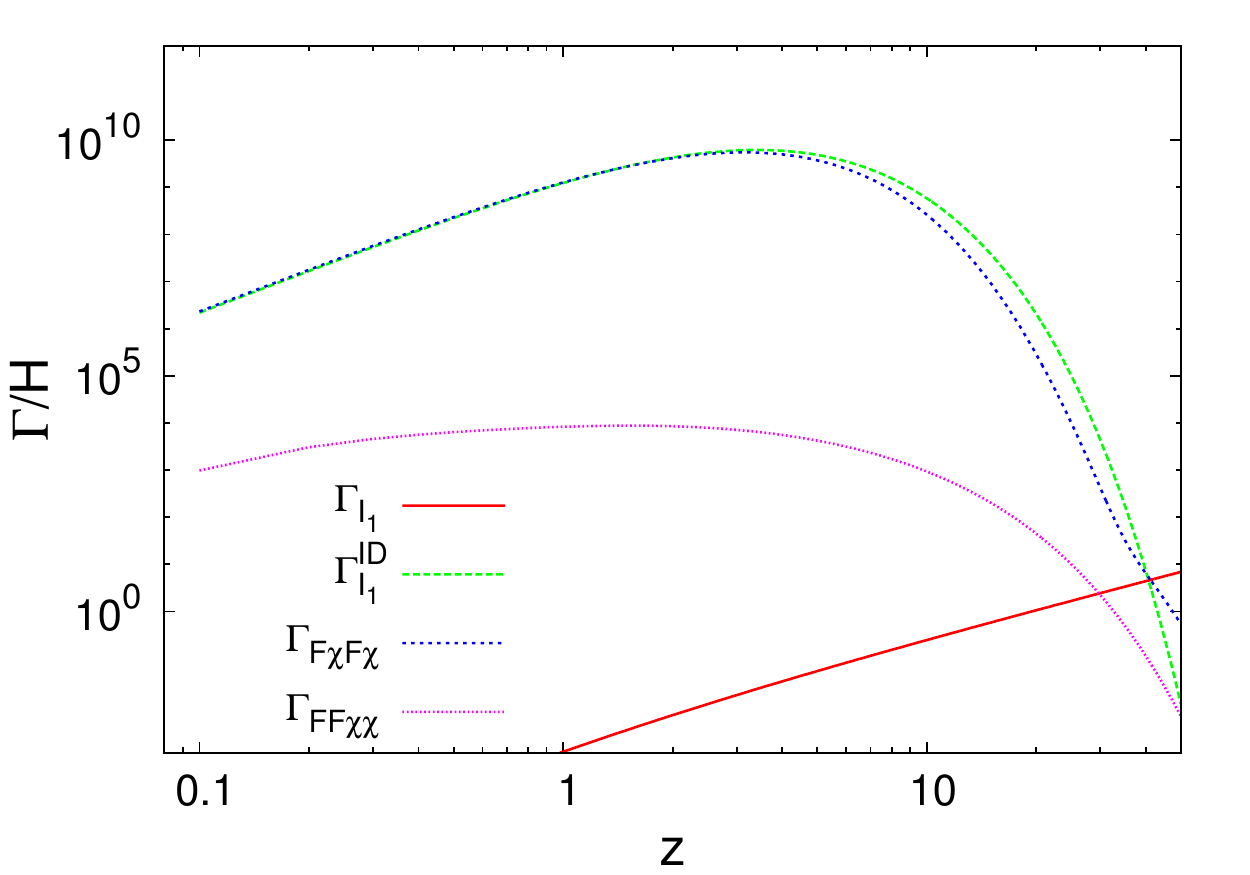}
\vspace{0.5cm}
\caption{Left panel: Evolution of $Y_B$, $Y_{N_1}$ and $Y_{X_1}$ where
 the horizontal black dashed line represents the required value of the baryon
 number asymmetry. Right panel: Reaction rates
 $\Gamma/H$ of the processes that have crucial effects for the
 baryon number asymmetry. } 
\label{resulep}
\end{figure}

\begin{figure}[t]
\includegraphics[scale=0.7]{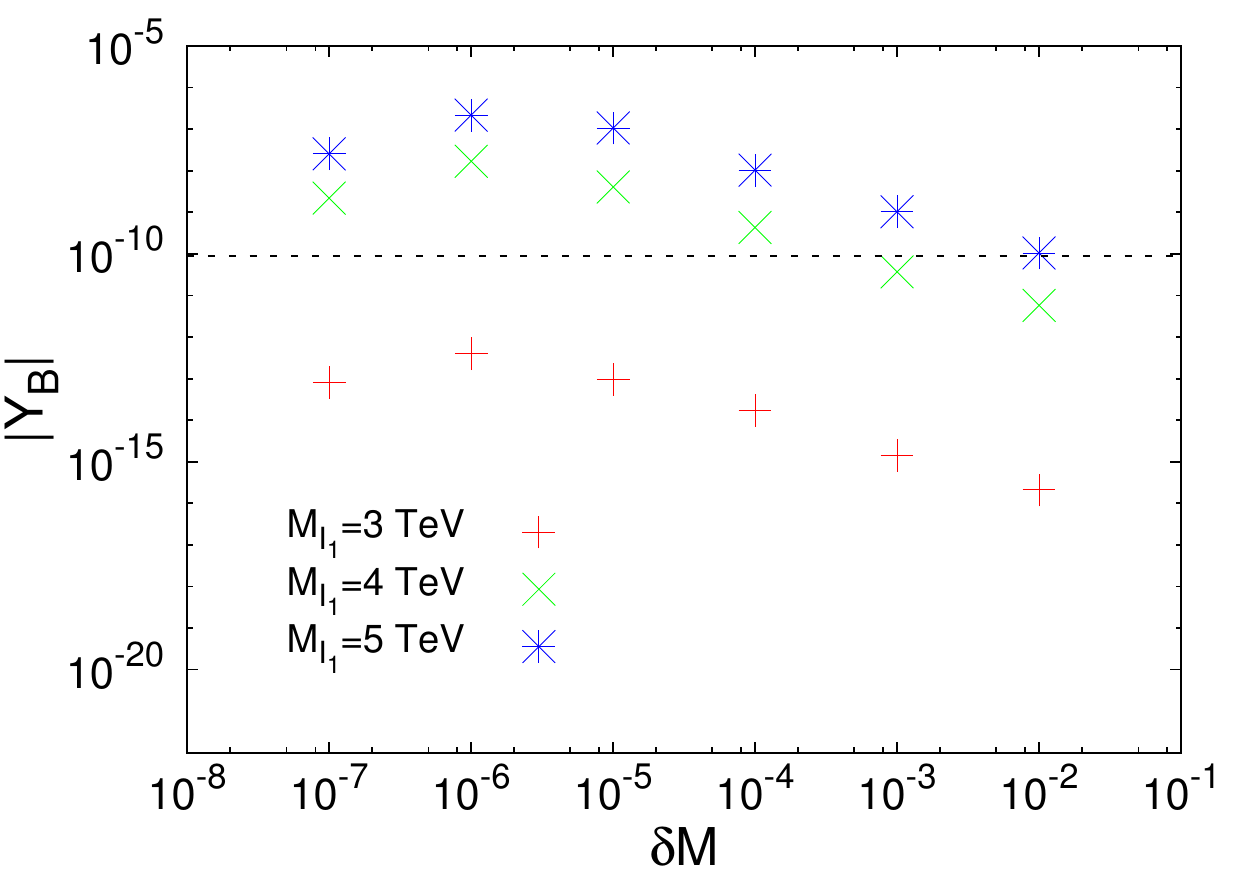}
\vspace{0.5cm}
\caption{$\delta M$ dependence of the generated baryon number asymmetry
 $Y_B$ with the different value of $M_{I_1}$. The value of $M_{I_1}$ is
 fixed to 3 TeV, 4 TeV and 5 TeV respectively.} 
\label{resolep}
\end{figure}

The decay of the lightest Majorana fermions should be out of thermal
equilibrium so that the lepton number asymmetry can be generated through
their decays. If we express the Hubble parameter as $H$, this condition
is given by $H>\Gamma_{I_1}$ at $T\sim M_{I_1}$. Since we assume that
each of the Majorana fermion mass is $M_{I_1}=5$ TeV,
$M_{I_2}=M_{I_1}(1+\delta M)$ and $M_{I_3}=6$ TeV, the Yukawa couplings of
$N_1$ and $X_1$ should be ${\cal O}(10^{-8})$. On the other hand, the
rest of the Yukawa couplings should be ${\cal O}(10^{-2})$ in order to
generate the appropriate neutrino masses. Here we set Yukawa
couplings in Eq.~(\ref{Ycoup}) to be $y_{I_1\chi}=1.5\times 10^{-8}$ and
$y_{I_{2,3}\chi}=10^{-2}$ and $\theta_i=\frac{\pi}{4}$. We show the
result for $\delta M=10^{-3}$ in Fig.~\ref{resulep} as an example, and also
the generated baryon number asymmetry for each value of $\delta M$ and
$M_{I_1}$ in Fig.~\ref{resolep}. Through this analysis, the masses of $F$,
$\chi$ and $\chi'$ are fixed to be $M_F=1.5$ TeV and $M_\chi=M_{\chi'}=
1.2$ TeV respectively. These parameter set satisfies the condition for
the DM phenomenology we discussed in the previous section. 

From the left panel in Fig.~\ref{resulep}, we can see that the required
baryon number asymmetry $Y_B$ can be obtained in this model. In the
right panel, we plot the behavior of the relevant reaction rate for each
process. This panel shows that the reaction rates of the inverse decay
process and the lepton number-violating process induced by the s-channel
$I_i$ exchange are quite large for a long time. Thus the baryon number
asymmetry cannot be generated quickly until rather a late period. After $T\sim
M_{I_1}$, the generated baryon number asymmetry gradually increases and
then the required value can be realized. This is because these processes
are suppressed by the Boltzmann factor.

We show the relation between the generated baryon number asymmetry and
the mass degeneracy of Majorana fermions in Fig.~\ref{resolep}.  
Notice here that the generated baryon number asymmetry is always smaller than the required value in the case $M_{I_1}\sim 3$ TeV. In this model, we can realize the large Yukawa couplings to explain the small neutrino mass due to the two-loop effects and then $\Gamma_{I_2}$ becomes larger compared to tree and one-loop neutrino mass models. Thus, the required mass degeneracy can be milder. However, the large Yukawa couplings cause the large washout effects.
Since the Boltzmann suppression does not work well in the case of small $M_{I_1}$, 
the large washout effects remain
 until quite a late period compared to the heavier cases. Thus the most
 of the generated baryon number asymmetry is washed out. 
We find that the observed baryon number asymmetry can be generated
when the mass degeneracy is $\delta M=(10^{-3}-10^{-2})$ for
$M_{I_1}=4-5$ TeV, as can be seen in Fig.~\ref{resolep}. As we mentioned
above, the magnitude of this mass degeneracy is quite milder than the canonical seesaw case for each
value of $M_{I_1}$ due to the loop effects.\footnote{The typical scale of $\delta M$ is
$10^{-10}-10^{-8}$. It is also worth mentioning that the Ma model has an
solution for $\delta M\le10^{-6.5}$~\cite{Kashiwase:2013uy}.}

\section{Conclusions}
We have studied a two-loop induced radiative neutrino model at TeV scale
with a $U(1)$ global symmetry, in which two types of DM candidates
(the lightest one of fermion or scalar) can be involved. 
The loop-induced neutrino masses have been evaluated appropriately and phenomenology
of DM has also been discussed. 
The fermionic DM candidate is disfavored if we reproduce the measured neutrino masses and take into account the constraint from LFV with the
same order of all the elements of $y_\eta$. Then we have found that 
the scalar can be a good DM candidate which satisfies the observed
relic density and the DM direct detection bound. 
We also found that the direct detection rate of DM is controlled by the
coupling $\lambda_{\Phi\eta}$ and some parameter region can be testable by the
next future direct detection experiment XENON1T.

We have discussed baryon number asymmetry through the resonant
leptogenesis with multi-sources scenario, in which the large Yukawa
couplings
(that is required to compensate the tiny neutrino masses at two-loop level)
make the large $CP$ asymmetry, but also
cause the large washout effects. We have shown that the required baryon
number asymmetry can be obtained for the parameter region {\it i.e.}, $\delta M =(10^{-3}-10^{-2})$ for $M_{I_1}=4-5$ TeV,
where $I\equiv N$ or $X$. In this case, the lightest Majorana fermions
should satisfy
$M_{I_1}\gtrsim3$ TeV to suppress the washout effects by the Boltzmann
factor. 
For larger $M_{I_1}$, the required magnitude of the mass degeneracy is rather
milder than the canonical seesaw case even at TeV scale models.  

\vspace{0.5cm}
\section*{Acknowledgments}
S.~K. is supported by
Grant-in-Aid for JSPS fellows (Grant No. 26 5862).
H.~O. expresses his sincere gratitude toward all the KIAS members,
Korean cordial persons, foods, culture, weather, and all the other
things. 
Y.~O. is supported by the Korea Neutrino Research Center which is
established by the National Research Foundation of Korea(NRF) grant
funded by the Korea government(MSIP) (No. 2009-0083526).
T.~T. acknowledges support from P2IO Excellence Laboratory. 


\begin{thebibliography}{99}


\bibitem{Ma:2006km} 
  E.~Ma,
  Phys.\ Rev.\ D {\bf 73}, 077301 (2006)
  [hep-ph/0601225].

\bibitem{Aoki:2013gzs} 
  M.~Aoki, J.~Kubo and H.~Takano,
  Phys.\ Rev.\ D {\bf 87}, no. 11, 116001 (2013)
  [arXiv:1302.3936 [hep-ph]].

\bibitem{Dasgupta:2013cwa} 
  B.~Dasgupta, E.~Ma and K.~Tsumura,
  Phys.\ Rev.\ D {\bf 89}, no. 4, 041702 (2014)
  [arXiv:1308.4138 [hep-ph]].

\bibitem{Krauss:2002px}
  L.~M.~Krauss, S.~Nasri and M.~Trodden,
  Phys.\ Rev.\  D {\bf 67}, 085002 (2003)
  [arXiv:hep-ph/0210389].

\bibitem{Aoki:2008av}
  M.~Aoki, S.~Kanemura and O.~Seto,
  Phys.\ Rev.\ Lett.\  {\bf 102}, 051805 (2009)
  [arXiv:0807.0361].

\bibitem{Baek:2012ub} 
  S.~Baek, P.~Ko, H.~Okada and E.~Senaha,
  JHEP {\bf 1409}, 153 (2014)
  [arXiv:1209.1685 [hep-ph]].

\bibitem{Schmidt:2012yg} 
  D.~Schmidt, T.~Schwetz and T.~Toma,
  Phys.\ Rev.\ D {\bf 85}, 073009 (2012)
  [arXiv:1201.0906 [hep-ph]].

\bibitem{Bouchand:2012dx} 
  R.~Bouchand and A.~Merle,
  JHEP {\bf 1207}, 084 (2012)
  [arXiv:1205.0008 [hep-ph]].


\bibitem{Aoki:2011he} 
  M.~Aoki, J.~Kubo, T.~Okawa and H.~Takano,
  Phys.\ Lett.\ B {\bf 707}, 107 (2012)
  [arXiv:1110.5403 [hep-ph]].


\bibitem{Farzan:2012sa} 
  Y.~Farzan and E.~Ma,
  Phys.\ Rev.\ D {\bf 86}, 033007 (2012)
  [arXiv:1204.4890 [hep-ph]].

\bibitem{Bonnet:2012kz} 
  F.~Bonnet, M.~Hirsch, T.~Ota and W.~Winter,
  JHEP {\bf 1207}, 153 (2012)
  [arXiv:1204.5862 [hep-ph]].

\bibitem{Kumericki:2012bf} 
  K.~Kumericki, I.~Picek and B.~Radovcic,
  JHEP {\bf 1207}, 039 (2012)
  [arXiv:1204.6597 [hep-ph]].

\bibitem{Kumericki:2012bh} 
  K.~Kumericki, I.~Picek and B.~Radovcic,
  Phys.\ Rev.\ D {\bf 86}, 013006 (2012)
  [arXiv:1204.6599 [hep-ph]].

\bibitem{Ma:2012if} 
  E.~Ma,
  Phys.\ Lett.\ B {\bf 717}, 235 (2012)
  [arXiv:1206.1812 [hep-ph]].

\bibitem{Gil:2012ya} 
  G.~Gil, P.~Chankowski and M.~Krawczyk,
  Phys.\ Lett.\ B {\bf 717}, 396 (2012)
  [arXiv:1207.0084 [hep-ph]].

\bibitem{Okada:2012np} 
  H.~Okada and T.~Toma,
  Phys.\ Rev.\ D {\bf 86}, 033011 (2012)
  arXiv:1207.0864 [hep-ph].

\bibitem{Hehn:2012kz} 
  D.~Hehn and A.~Ibarra,
  Phys.\ Lett.\ B {\bf 718}, 988 (2013)
  [arXiv:1208.3162 [hep-ph]].

\bibitem{Dev:2012sg} 
  P.~S.~B.~Dev and A.~Pilaftsis,
  Phys.\ Rev.\ D {\bf 86}, 113001 (2012)
  [arXiv:1209.4051 [hep-ph]].

\bibitem{Kajiyama:2012xg} 
  Y.~Kajiyama, H.~Okada and T.~Toma,
  Eur.\ Phys.\ J.\ C {\bf 73}, no. 3, 2381 (2013)
  [arXiv:1210.2305 [hep-ph]].

\bibitem{Okada:2012sp} 
  H.~Okada,
  arXiv:1212.0492 [hep-ph].

\bibitem{Aoki:2010ib} 
  M.~Aoki, S.~Kanemura, T.~Shindou and K.~Yagyu,
  JHEP {\bf 1007}, 084 (2010)
  [Erratum-ibid.\  {\bf 1011}, 049 (2010)]
  [arXiv:1005.5159 [hep-ph]].

\bibitem{Kanemura:2011vm} 
  S.~Kanemura, O.~Seto and T.~Shimomura,
  Phys.\ Rev.\ D {\bf 84}, 016004 (2011)
  [arXiv:1101.5713 [hep-ph]].

\bibitem{Lindner:2011it} 
  M.~Lindner, D.~Schmidt and T.~Schwetz,
  Phys.\ Lett.\ B {\bf 705}, 324 (2011)
  [arXiv:1105.4626 [hep-ph]].

\bibitem{Kanemura:2011mw} 
  S.~Kanemura, T.~Nabeshima and H.~Sugiyama,
  Phys.\ Rev.\ D {\bf 85}, 033004 (2012)
  [arXiv:1111.0599 [hep-ph]].


\bibitem{Kanemura:2012rj} 
  S.~Kanemura and H.~Sugiyama,
  Phys.\ Rev.\ D {\bf 86}, 073006 (2012)
  [arXiv:1202.5231 [hep-ph]].

\bibitem{Gu:2007ug} 
  P.~-H.~Gu and U.~Sarkar,
  Phys.\ Rev.\ D {\bf 77}, 105031 (2008)
  [arXiv:0712.2933 [hep-ph]].

\bibitem{Gu:2008zf} 
  P.~-H.~Gu and U.~Sarkar,
  Phys.\ Rev.\ D {\bf 78}, 073012 (2008)
  [arXiv:0807.0270 [hep-ph]].

\bibitem{Gustafsson} 
M.~Gustafsson, J.~M.~No and M.~A.~Rivera,
  Phys.\ Rev.\ Lett.\  {\bf 110}, 211802 (2013)
arXiv:1212.4806 [hep-ph].

\bibitem{Kajiyama:2013zla} 
  Y.~Kajiyama, H.~Okada and K.~Yagyu,
  Nucl.\ Phys.\ B {\bf 874}, 198 (2013)
  [arXiv:1303.3463 [hep-ph]].

\bibitem{Kajiyama:2013rla} 
  Y.~Kajiyama, H.~Okada and T.~Toma,
  Phys.\ Rev.\ D {\bf 88}, 015029 (2013)
  [arXiv:1303.7356].
 

\bibitem{Hernandez:2013dta} 
  A.~E.~Carcamo Hernandez, I.~d.~M.~Varzielas, S.~G.~Kovalenko, H.~P\"as and I.~Schmidt,
  Phys.\ Rev.\ D {\bf 88}, 076014 (2013)
  [arXiv:1307.6499 [hep-ph]].

\bibitem{Hernandez:2013hea} 
  A.~E.~Carcamo Hernandez, RMartinez and F.~Ochoa,
  arXiv:1309.6567 [hep-ph].

\bibitem{McDonald:2013hsa} 
  K.~L.~McDonald,
  JHEP {\bf 1311}, 131 (2013)
  [arXiv:1310.0609 [hep-ph]].


\bibitem{Okada:2013iba} 
  H.~Okada and K.~Yagyu,
  Phys.\ Rev.\ D {\bf 89}, no. 5, 053008 (2014)
  [arXiv:1311.4360 [hep-ph]].

\bibitem{Baek:2013fsa} 
  S.~Baek, H.~Okada and T.~Toma,
  JCAP {\bf 1406}, 027 (2014)
  [arXiv:1312.3761 [hep-ph]].

\bibitem{Ma:2014cfa} 
  E.~Ma,
  Phys.\ Lett.\ B {\bf 732}, 167 (2014)
  [arXiv:1401.3284 [hep-ph]].

\bibitem{Baek:2014awa} 
  S.~Baek, H.~Okada and T.~Toma,
  Phys.\ Lett.\ B {\bf 732}, 85 (2014)
  [arXiv:1401.6921 [hep-ph]].

\bibitem{Ahriche:2014xra} 
  A.~Ahriche, S.~Nasri and R.~Soualah,
  Phys.\ Rev.\ D {\bf 89}, no. 9, 095010 (2014)
  [arXiv:1403.5694 [hep-ph]].
  
\bibitem{Kanemura:2011jj}
 S.~Kanemura, T.~Nabeshima and H.~Sugiyama,
 Phys.\ Lett.\ B {\bf 703}, 66 (2011)
 [arXiv:1106.2480 [hep-ph]].

\bibitem{Kanemura:2013qva}
 S.~Kanemura, T.~Matsui and H.~Sugiyama,
Higgs Doublet Model,''
 Phys.\ Lett.\ B {\bf 727}, 151 (2013)
 [arXiv:1305.4521 [hep-ph]].
 
\bibitem{Okada:2014nsa} 
  H.~Okada and K.~Yagyu,
  Phys.\ Rev.\ D {\bf 90}, no. 3, 035019 (2014)
  [arXiv:1405.2368 [hep-ph]].

  

\bibitem{Kanemura:2014rpa} 
  S.~Kanemura, T.~Matsui and H.~Sugiyama,
  Phys.\ Rev.\ D {\bf 90}, 013001 (2014)
  [arXiv:1405.1935 [hep-ph]].
  

\bibitem{Chen:2014ska} 
  C.~S.~Chen, K.~L.~McDonald and S.~Nasri,
  Phys.\ Lett.\ B {\bf 734}, 388 (2014)
  [arXiv:1404.6033 [hep-ph]].

\bibitem{Ahriche:2014oda} 
  A.~Ahriche, K.~L.~McDonald and S.~Nasri,
  JHEP {\bf 1410}, 167 (2014)
  [arXiv:1404.5917 [hep-ph]].
  
\bibitem{Okada:2014vla} 
  H.~Okada,
  arXiv:1404.0280 [hep-ph].

\bibitem{Ahriche:2014cda} 
  A.~Ahriche, C.~S.~Chen, K.~L.~McDonald and S.~Nasri,
  Phys.\ Rev.\ D {\bf 90}, 015024 (2014)
  [arXiv:1404.2696 [hep-ph]].

\bibitem{Aoki:2014cja} 
  M.~Aoki and T.~Toma,
  JCAP {\bf 1409}, 016 (2014)
  [arXiv:1405.5870 [hep-ph]].

\bibitem{Lindner:2014oea} 
  M.~Lindner, S.~Schmidt and J.~Smirnov,
  JHEP {\bf 1410}, 177 (2014)
  [arXiv:1405.6204 [hep-ph]].
  
  
\bibitem{Ahn:2012cg} 
  Y.~H.~Ahn and H.~Okada,
  Phys.\ Rev.\ D {\bf 85}, 073010 (2012)
  [arXiv:1201.4436 [hep-ph]].

\bibitem{Ma:2012ez} 
  E.~Ma, A.~Natale and A.~Rashed,
  Int.\ J.\ Mod.\ Phys.\ A {\bf 27}, 1250134 (2012)
  [arXiv:1206.1570 [hep-ph]].
  
\bibitem{Kajiyama:2013lja} 
  Y.~Kajiyama, H.~Okada and K.~Yagyu,
  JHEP {\bf 10}, 196 (2013)
  arXiv:1307.0480 [hep-ph].
  
\bibitem{Kajiyama:2013sza} 
  Y.~Kajiyama, H.~Okada and K.~Yagyu,
  Nucl.\ Phys.\ B {\bf 887}, 358 (2014)
  [arXiv:1309.6234 [hep-ph]].

\bibitem{Ma:2013mga} 
  E.~Ma,
  Phys.\ Rev.\ Lett.\  {\bf 112}, 091801 (2014)
  [arXiv:1311.3213 [hep-ph]].
  
\bibitem{Ma:2014eka} 
  E.~Ma and A.~Natale,
  Phys.\ Lett.\ B {\bf 734}, 403 (2014)
  [arXiv:1403.6772 [hep-ph]].
   
 
 
\bibitem{Kanemura:2015mxa} 
  S.~Kanemura, M.~Kikuchi and K.~Yagyu,
  Nucl.\ Phys.\ B {\bf 896}, 80 (2015)
  [arXiv:1502.07716 [hep-ph]].
\bibitem{Bahrami:2015mwa} 
  S.~Bahrami and M.~Frank,
  Phys.\ Rev.\ D {\bf 91}, 075003 (2015)
  [arXiv:1502.02680 [hep-ph]].
\bibitem{Baek:2015mna} 
  S.~Baek, H.~Okada and K.~Yagyu,
  JHEP {\bf 1504}, 049 (2015)
  [arXiv:1501.01530 [hep-ph]].
\bibitem{Hatanaka:2014tba} 
  H.~Hatanaka, K.~Nishiwaki, H.~Okada and Y.~Orikasa,
  Nucl.\ Phys.\ B {\bf 894}, 268 (2015)
  [arXiv:1412.8664 [hep-ph]].
\bibitem{Okada:2014nea} 
  H.~Okada and Y.~Orikasa,
  arXiv:1412.3616 [hep-ph].
\bibitem{Sierra:2014rxa} 
  D.~Aristizabal Sierra, A.~Degee, L.~Dorame and M.~Hirsch,
  JHEP {\bf 1503}, 040 (2015)
  [arXiv:1411.7038 [hep-ph]].
\bibitem{Okada:2014qsa} 
  H.~Okada, T.~Toma and K.~Yagyu,
  Phys.\ Rev.\ D {\bf 90}, no. 9, 095005 (2014)
  [arXiv:1408.0961 [hep-ph]].
	


\bibitem{Hernandez:2015tna} 
  A.~E.~Carcamo Hernandez and R.~Martinez,
  arXiv:1501.05937 [hep-ph].
  
  
\bibitem{Hernandez:2015cra} 
  A.~E.~C.~Hernandez and R.~Martinez,
  arXiv:1501.07261 [hep-ph].
  
  
\bibitem{Culjak:2015qja} 
  P.~Culjak, K.~Kumericki and I.~Picek,
  Phys.\ Lett.\ B {\bf 744}, 237 (2015)
  [arXiv:1502.07887 [hep-ph]].

\bibitem{Humbert:2015epa} 
  P.~Humbert, M.~Lindner and J.~Smirnov,
  JHEP {\bf 1506}, 035 (2015)
  [arXiv:1503.03066 [hep-ph]].
  
\bibitem{Okada:2015nga} 
  H.~Okada,
  arXiv:1503.04557 [hep-ph].
  
\bibitem{Geng:2015sza} 
  C.~Q.~Geng and L.~H.~Tsai,
  arXiv:1503.06987 [hep-ph].
  
\bibitem{Okada:2015bxa} 
  H.~Okada, N.~Okada and Y.~Orikasa,
  arXiv:1504.01204 [hep-ph].
  
\bibitem{Geng:2015coa} 
  C.~Q.~Geng, D.~Huang and L.~H.~Tsai,
  Phys.\ Lett.\ B {\bf 745}, 56 (2015)
  [arXiv:1504.05468 [hep-ph]].

\bibitem{Ahriche:2015wha} 
  A.~Ahriche, K.~L.~McDonald, S.~Nasri and T.~Toma,
  Phys.\ Lett.\ B {\bf 746}, 430 (2015)
  [arXiv:1504.05755 [hep-ph]].
    
\bibitem{Ahriche:2015lba} 
  A.~Ahriche, K.~L.~McDonald and S.~Nasri,
  arXiv:1504.06759 [hep-ph].
  
  


\bibitem{Okada:2014oda}
Hiroshi Okada, Yuta Orikasa,
Phys.Rev. \textbf{D90}, 075023 (2014),
arXiv:1407.2543 [hep-ph].









\bibitem{Kashiwase:2013uy} 
  S.~Kashiwase and D.~Suematsu,
  Eur.\ Phys.\ J.\ C {\bf 73}, 2484 (2013)
  [arXiv:1301.2087 [hep-ph]].


\bibitem{Appelquist:2002mw} 
  T.~Appelquist, B.~A.~Dobrescu and A.~R.~Hopper,
  Phys.\ Rev.\ D {\bf 68}, 035012 (2003)
  [hep-ph/0212073].

\bibitem{Batra:2005rh} 
  P.~Batra, B.~A.~Dobrescu and D.~Spivak,
  J.\ Math.\ Phys.\  {\bf 47}, 082301 (2006)
  [hep-ph/0510181].



\bibitem{Maki:1962mu} 
  Z.~Maki, M.~Nakagawa and S.~Sakata,
  Prog.\ Theor.\ Phys.\  {\bf 28}, 870 (1962).

\bibitem{Forero:2014bxa} 
  D.~V.~Forero, M.~Tortola and J.~W.~F.~Valle,
  Phys.\ Rev.\ D {\bf 90}, no. 9, 093006 (2014)
  [arXiv:1405.7540 [hep-ph]].

\bibitem{Borowka:2015mxa} 
  S.~Borowka, G.~Heinrich, S.~P.~Jones, M.~Kerner, J.~Schlenk and
  T.~Zirke,
  Comput.\ Phys.\ Commun.\  {\bf 196}, 470 (2015)
  [arXiv:1502.06595 [hep-ph]].

\bibitem{Toma:2013zsa} 
  T.~Toma and A.~Vicente,
  JHEP {\bf 1401}, 160 (2014)
  [arXiv:1312.2840 [hep-ph]].



\bibitem{Adam:2013mnn} 
  J.~Adam {\it et al.}  [MEG Collaboration],
  Phys.\ Rev.\ Lett.\  {\bf 110}, 201801 (2013)
  [arXiv:1303.0754 [hep-ex]].

\bibitem{Cheung:2013kla} 
  K.~Cheung, J.~S.~Lee and P.~Y.~Tseng,
  JHEP {\bf 1305}, 134 (2013)
  [arXiv:1302.3794 [hep-ph]].


\bibitem{Robens:2015gla} 
  T.~Robens and T.~Stefaniak,
  Eur.\ Phys.\ J.\ C {\bf 75}, no. 3, 104 (2015)
  [arXiv:1501.02234 [hep-ph]].

\bibitem{Keung:2013mfa} 
  W.~Y.~Keung, K.~W.~Ng, H.~Tu and T.~C.~Yuan,
  Phys.\ Rev.\ D {\bf 90}, no. 7, 075014 (2014)
  doi:10.1103/PhysRevD.90.075014
  [arXiv:1312.3488 [hep-ph]].

  
\bibitem{Ibe:2011hq} 
  M.~Ibe, S.~Matsumoto and T.~T.~Yanagida,
  Phys.\ Lett.\ B {\bf 708}, 112 (2012)
  [arXiv:1110.5452 [hep-ph]].







\bibitem{Belanger:2013oya} 
  G.~Belanger, F.~Boudjema, A.~Pukhov and A.~Semenov,
  Comput.\ Phys.\ Commun.\  {\bf 185}, 960 (2014)
  [arXiv:1305.0237 [hep-ph]].


\bibitem{Ade:2015xua} 
  P.~A.~R.~Ade {\it et al.} [Planck Collaboration],
  arXiv:1502.01589 [astro-ph.CO].

\bibitem{Akerib:2013tjd} 
  D.~S.~Akerib {\it et al.} [LUX Collaboration],
  Phys.\ Rev.\ Lett.\  {\bf 112}, 091303 (2014)
  [arXiv:1310.8214 [astro-ph.CO]].

\bibitem{Aprile:2012zx} 
  E.~Aprile [XENON1T Collaboration],
  Springer Proc.\ Phys.\  {\bf 148}, 93 (2013)
  [arXiv:1206.6288 [astro-ph.IM]].


\bibitem{Hambye:2009pw} 
  T.~Hambye, F.-S.~Ling, L.~Lopez Honorez and J.~Rocher,
  JHEP {\bf 0907}, 090 (2009)
  Erratum: [JHEP {\bf 1005}, 066 (2010)]
  [arXiv:0903.4010 [hep-ph]].



\bibitem{Fukugita:1986hr} 
  M.~Fukugita and T.~Yanagida,
  Phys.\ Lett.\ B {\bf 174}, 45 (1986).
\bibitem{Flanz:1994yx} 
  M.~Flanz, E.~A.~Paschos and U.~Sarkar,
  Phys.\ Lett.\ B {\bf 345}, 248 (1995)
  [Erratum-ibid.\ B {\bf 382}, 447 (1996)]
  [hep-ph/9411366].

\bibitem{Covi:1996wh} 
  L.~Covi, E.~Roulet and F.~Vissani,
  Phys.\ Lett.\ B {\bf 384}, 169 (1996)
  [hep-ph/9605319].

\bibitem{Pilaftsis:1997jf} 
  A.~Pilaftsis,
  Phys.\ Rev.\ D {\bf 56}, 5431 (1997)
  [hep-ph/9707235].

\bibitem{Akhmedov:2003dg} 
  E.~K.~Akhmedov, M.~Frigerio and A.~Y.~Smirnov,
  JHEP {\bf 0309}, 021 (2003)
  [hep-ph/0305322].

\bibitem{Albright:2003xb} 
  C.~H.~Albright and S.~M.~Barr,
  Phys.\ Rev.\ D {\bf 69}, 073010 (2004)
  [hep-ph/0312224].

\bibitem{Hambye:2004jf} 
  T.~Hambye, J.~March-Russell and S.~M.~West,
  JHEP {\bf 0407}, 070 (2004)
  [hep-ph/0403183].

\bibitem{Pilaftsis:2003gt} 
  A.~Pilaftsis and T.~E.~J.~Underwood,
  Nucl.\ Phys.\ B {\bf 692}, 303 (2004)
  [hep-ph/0309342].

\bibitem{Pilaftsis:2005rv} 
  A.~Pilaftsis and T.~E.~J.~Underwood,
  Phys.\ Rev.\ D {\bf 72}, 113001 (2005)
  [hep-ph/0506107].
  
\bibitem{Dev:2014laa} 
  P.~S.~Bhupal Dev, P.~Millington, A.~Pilaftsis and D.~Teresi,
  Nucl.\ Phys.\ B {\bf 886}, 569 (2014)
  [arXiv:1404.1003 [hep-ph]].

\bibitem{Kashiwase:2012xd} 
  S.~Kashiwase and D.~Suematsu,
  Phys.\ Rev.\ D {\bf 86}, 053001 (2012)
  [arXiv:1207.2594 [hep-ph]].





\bibitem{Kolb:1979qa} 
  E.~W.~Kolb and S.~Wolfram,
  Nucl.\ Phys.\ B {\bf 172}, 224 (1980)
  [Erratum-ibid.\ B {\bf 195}, 542 (1982)].


\bibitem{Luty:1992un} 
  M.~A.~Luty,
  Phys.\ Rev.\ D {\bf 45}, 455 (1992).

\bibitem{Plumacher:1997ru} 
  M.~Plumacher,
  Nucl.\ Phys.\ B {\bf 530}, 207 (1998)
  [hep-ph/9704231].

\end{thebibliography}
\end{document}